\documentclass[sn-mathphys-num]{sn-jnl}

\usepackage{graphicx}%
\usepackage{multirow}%
\usepackage{amsmath,amssymb,amsfonts}%
\usepackage{amsthm}%
\usepackage{mathrsfs}%
\usepackage[title]{appendix}%
\usepackage{xcolor}%
\usepackage{textcomp}%
\usepackage{manyfoot}%
\usepackage{booktabs}%
\usepackage{algorithm}%
\usepackage{algorithmicx}%
\usepackage{algpseudocode}%
\usepackage{listings}%
\usepackage{subcaption}%

\usepackage{xr}
\makeatletter
\newcommand*{\addFileDependency}[1]{
  \typeout{(#1)}
  \@addtofilelist{#1}
  \IfFileExists{#1}{}{\typeout{No file #1.}}
}
\makeatother

\newcommand*{\myexternaldocument}[1]{%
    \externaldocument{#1}%
    \addFileDependency{#1.tex}%
    \addFileDependency{#1.aux}%
}
\myexternaldocument{suppl}

\begin{document}

\title{Using Stochastic Block Models for Community Detection: The issue of edge-connectivity}



\author[1]{\fnm{The-Anh} \sur{Vu-Le}}\email{vltanh@illinois.edu}
\equalcont{These authors contributed equally to this work.}
\author[1]{\fnm{Minhyuk} \sur{Park}}\email{minhyuk2@illinois.edu}
\equalcont{These authors contributed equally to this work.}
\author[1]{\fnm{Ian} \sur{Chen}}\email{ianchen3@illinois.edu}



\author[1]{\fnm{George} \sur{Chacko}}\email{chackoge@illinois.edu}

\author*[1]{\fnm{Tandy} \sur{Warnow}}\email{warnow@illinois.edu}

\affil[1]{\orgdiv{Siebel School of Computing and Data Science}, \orgname{University of Illinois Urbana-Champaign}, \orgaddress{\street{201 N. Goodwin Avenue}, \city{Urbana}, \postcode{61801}, \state{IL}, \country{USA}}}





 \abstract{A relevant, sometimes overlooked, quality criterion for communities in graphs is that they should be  \emph{well-connected} in addition to being edge-dense. Prior work has shown that  leading community detection methods can produce poorly-connected communities, and some even produce internally disconnected communities. A recent study by Park {\em et al.} in Complex Networks and their Applications 2024  showed that this problem is evident in clusterings from three Stochastic Block Models (SBMs) in graph-tool, a popular software package. To address this issue, Park et al.~presented a simple technique, Well-Connected Clusters (WCC), that repeatedly finds and removes small edge cuts of size at most $\log_{10}n$ in clusters, where $n$ is the number of nodes in the cluster, and showed that treatment of graph-tool SBM clusterings with WCC improves accuracy. Here we examine the question of cluster connectivity for clusterings computed using other SBM software or  nested SBMs within graph-tool. Our study, using a wide range of real-world and synthetic networks, shows that all tested SBM clustering methods produce communities that are disconnected, and that graph-tool improves on PySBM.  
 We provide insight into why graph-tool degree-corrected SBM clustering produces disconnected clusters by examining the description length formula it uses, and explore the impact of modifications to the description length formula. Finally, we show that WCC provides an improvement in accuracy for both flat and nested SBMs and establish that it scales to networks with millions of nodes.
 }




\maketitle

\section*{Introduction}
\label{sec:intro}


Community detection, also known as graph clustering, is the problem of taking a graph and partitioning the vertices into disjoint subsets so that each set has properties indicative of being a community.
Important community properties include
edge-density, separability from the rest of the graph, and well-connectedness. To be well-connected, each community should not have a small edge cut, i.e., it should not be disconnected by the deletion of a small number of edges \cite{kannan2004clusterings,leiden,Yang2013}.

Despite the importance of edge connectivity, some clustering methods produce poorly-connected clusters, and some even produce internally disconnected clusters.
As an example, the Louvain algorithm \cite{louvain}, which is often used to find modularity-based clusterings, has been shown to produce disconnected communities \cite{leiden}.
Because of this problem with Louvain, the Leiden algorithm \cite{leiden} was developed, which is guaranteed to produce connected clusters.

However, it is clear that other clustering methods also have the potential to produce disconnected clusters. Using a mild standard for ``well-connected", by which a cluster of $n$ nodes is considered well-connected only when the size of its minimum edge cut is greater than $\log_{10}(n)$, Park and colleagues \cite{cm-journal} found that the Leiden algorithm \cite{leiden-code,leiden} optimizing modularity or the Constant Potts Model (CPM), Infomap \cite{infomap}, Iterative-K-core Clustering (IKC) \cite{ikc} and Markov Clustering (MCL) \cite{mcl} all
produced poorly-connected clusters under some conditions.   
Also in \cite{cm-journal}, the authors presented the Connectivity Modifier (CM), a post-processing technique to improve the edge-connectivity of clusters and demonstrated improved clustering accuracy on a selection of synthetic networks.

A recent study by Park and colleagues \cite{park2024improved-conf-proceedings} evaluated the connectivity of clusters produced using some Stochastic Block Models (SBM) implemented in the graph-tool software \cite{graph-tool}. 
They found that graph-tool clustering using three ``flat" models--degree-corrected \cite{dcsbm} (DC-Flat), non degree-corrected \cite{holland1983sbm} (NDC-Flat), and planted partition \cite{ppsbm} (PP-Flat)--frequently produced disconnected communities. Post-processing by CM had a variable impact, sometimes improving accuracy and sometimes decreasing accuracy. 

Park and colleagues \cite{park2024improved-conf-proceedings} proposed two other techniques: Connected Components (CC) and Well-Connected Clusters (WCC).  
The CC technique replaces every cluster with its connected components, and so is a very simple technique. WCC processes each cluster independently, iteratively finding and removing small edge cuts (if they exist) until the cluster is well-connected (according to a user-specified bound). 
They observed that postprocessing clusterings by CC or WCC often improved accuracy on synthetic Lancichinetti-Fortunato-Radicchi (LFR) benchmark networks \cite{lfr} for clusterings produced by the best graph-tool's SBMs (defined by minimum description length) from DC-Flat, NDC-Flat, and PP-Flat. 
Thus, the authors of \cite{park2024improved-conf-proceedings} concluded that it is beneficial to use WCC with clusterings produced using graph-tool's flat models.
They also provided an explanation based on the description length for why flat degree-corrected SBMs produce disconnected clusters, especially on large networks.

However, the study reported in \cite{park2024improved-conf-proceedings} was limited in several ways. 
Most importantly, the study was limited to clusterings produced by graph-tool using its flat models (degree-corrected, non-degree-corrected, and planted partition), and did not examine hierarchical (nested) models available within graph-tool, which are expected to be robust to the SBM resolution limit \cite{pysbm,peixoto2019bayesian,zhang2020statistical,zhang2023realistic}, whereby the number of clusters that SBMs can detect has an upper bound that depends on the network size.  
Furthermore, alternative SBM software was not tested.  
Additional limitations include examining accuracy only with the   
Adjusted Rand Index (ARI) \cite{hubert1985comparing}, when additional accuracy criteria, such as Adjusted Mutual Information (AMI) \cite{vinh2009information-AMI}, are also relevant, plus using only a few LFR synthetic networks when other synthetic network generators could also have been considered for better reflections of real-world community structure.
The explanation based on the description length for why degree-corrected SBMs produce disconnected clusters was similarly limited, since other models were not examined.
Thus, although \cite{park2024improved-conf-proceedings} provided insight into SBM clustering and the benefit of using WCC to improve clustering accuracy, its limitations to a subset of the models and software are significant, and require further evaluation under additional conditions. 

Here, we report results from a study that provides a deeper investigation into the properties of clusterings produced using Stochastic Block Models, and the impact of using either CC or WCC for postprocessing.  
First, we establish that clustering using PySBM \cite{pysbm,pysbm-github} or using nested SBMs within graph-tool also produces disconnected clusters, indicating that this issue is not limited to just graph-tool's flat models.
We also establish that graph-tool is superior to PySBM in terms of finding good solutions to its optimization problems, i.e., minimizing the description length. 
Next, we show that nearly all clusters in SBM clusterings of bipartite networks are disconnected.
Based on these findings, we restrict the remainder of the study to the use of graph-tool (both flat and nested models) on non-bipartite networks.  
We then explore the impact of WCC and CC on clustering accuracy using three different accuracy criteria (NMI \cite{thomas2006elements} and AMI \cite{vinh2009information-AMI} in addition to ARI), for a large set of synthetic networks generated using the new synthetic network generator, EC-SBM \cite{vu2025ecsbm}, which is able to produce more realistic networks than LFR.  
These experiments establish that both CC and WCC improve the ability to recover high-quality clusters, but have reduced accuracy with respect to recovering very poor-quality clusters, such as large clusters that are very sparse, which we show are typically produced by modularity-based clustering.
We also extend the exploration of the description length formula and its impact on cluster connectivity.
Whereas \cite{park2024improved-conf-proceedings} provided an explanation based on the description length formula for why degree-corrected flat SBMs tend to be disconnected, the authors did not explore other models, and their evaluation was limited to one model condition. Here, we provide additional exploration of the impact of the modifications to the description length formula under a larger range of model conditions, and find cases where changes to the formula {\em do} show the potential to improve cluster connectivity.
Finally, \cite{park2024improved-conf-proceedings} did not explore computational performance, and so here we report runtime data for treatments of SBM on large real-world networks.

Overall, we present empirical results indicating that SBM clustering, using a variety of models and methods, tends to produce a modest number  
of communities, which tend to be sparse and are often internally disconnected. 
Our study demonstrates that SBMs can provide high overall accuracy {\em when} the networks have a very large fraction of the nodes in large sparse communities, and in these cases, applying WCC post-processing may reduce accuracy.
However, for networks where the network community structure has dense clusters, then applying WCC post-processing improves accuracy.

\section*{Materials and Methods} 
\label{sec:methods}

\subsection*{Real-world networks}

We used a set of $112$ real-world networks: $110$ from the Netzschleuder network catalogue \cite{netzschleuder} and $2$ other networks from \cite{cm-journal} (see Supplementary Materials, Section A 
 for the full list of networks).
The smallest of these networks (\texttt{dnc}) has $906$ nodes, and the largest (\texttt{CEN}) contains almost $14$ million nodes. 
The four largest networks, \texttt{livejournal}, \texttt{orkut}, \texttt{bitcoin}, and \texttt{CEN}, have at least three million nodes each.
Thus, $108$ of these networks are small-to-moderate in size (from $906$ to around $1.4$ millions nodes), and four are large (at least $3$ million nodes).
We use the four large networks in a final experiment evaluating computational performance, and the remaining ones for the other experiments.

Networks were obtained as lists of edges without directionality or weights and pre-processed to remove any self-loops or excessive duplicate edges when present; thus, each network was studied as undirected, unweighted, and simple.

\subsection*{Synthetic network generation}
 
We utilized the Edge-Connected SBM Network Generator (EC-SBM) \cite{vu2025ecsbm} as the synthetic network generator.
EC-SBM takes as input a network $N$ and some numeric parameters obtained from a clustering $\mathcal{C}$ of $N$,  and then computes a synthetic network modeled on the input, with the goal of reproducing as well as possible various network and clustering statistics.   
Specifically, EC-SBM is guaranteed to produce a synthetic network with the same number of clusters and sizes, and attempts to come close to other network statistics such as diameter, local and global clustering coefficient, and degree sequence.
It also aims to produce the same minimum edge-cut size as the given input clustering.   
EC-SBM uses graph-tool SBMs to produce some parts of the network, but modifies the network edges in order to improve the fit to the input parameters. 
Thus, for different clusterings of a given real-world network, EC-SBM produces different synthetic networks.  

As shown in \cite{vu2025ecsbm}, EC-SBM produces networks that better fit their input networks than graph-tool SBMs based on the same input, and are also better in this respect than other synthetic networks, such as LFR \cite{lfr} and RECCS \cite{anne2025reccs-journal}. 
Furthermore, the fidelity between the synthetic network and the real-world network depends on the input clustering. The highest fidelity was obtained when the input clustering was SBM+WCC, where SBM refers to the flat SBM model that had the lowest description length from graph-tool and the second highest fidelity was obtained using Leiden-Mod+CM as the input clustering \cite{vu2025ecsbm}.

Since we were evaluating the accuracy of clusterings using SBMs based on software other than graph-tool alone, and we also wished to evaluate the impact of WCC treatments, to avoid favorable results, we did not select EC-SBM networks based on input clusterings that were postprocessed by WCC or CM treatments. Therefore, we selected four clustering methods for use with EC-SBM network generation. 
We picked one that provided the best fit to the real-world network statistics of these clustering methods when given as input to EC-SBM:  Leiden-CPM(0.1).
We also picked one that had the worst fit (Leiden-Mod), and the remaining two that had intermediate fit  (Leiden-CPM(0.01) and SBM+CC), where SBM refers to the flat SBM model having the lowest description length.

Thus, we used EC-SBM to generate $296$ synthetic networks, each based on one of $4$ different clustering methods and one of the $74$ real-world non-bipartite networks described in the previous section. 
These networks range in size from around $1,000$ nodes (\texttt{dnc}) to slightly over $1.4$ million nodes (\texttt{hyves}).

\subsection*{Clustering using SBMs} 
\label{sec:methods-sbm}

For an input network, we generated SBM clusterings using three different approaches.
\begin{itemize}
    \item 
We used graph-tool \cite{graph-tool} to produce a ``chosen" SBM clustering, with the following protocol. 
First, we clustered the network using three different flat SBM models: DC-Flat, NDC-Flat, and PP-Flat. 
We then selected the clustering that achieved the lowest description length. 
This clustering is referred to as the 
``Chosen-Flat'' clustering of the input network. 
\item We used graph-tool \cite{graph-tool} to produce a ``chosen'' hierarchical (nested) SBM clustering, with the following protocol. First, we clustered the network using two different nested SBM models: the degree-corrected nested model (DC-Nested) and the non-degree-corrected nested model (NDC-Nested).  
We then selected the model with the lower description length. Finally, the bottom level (i.e., most refined) of the selected model was returned as clustering. 
The model (DC-Nested or NDC-Nested) having the lower description length is the 
``Chosen-Nested'' clustering of the input network. 
\item 
We used the PySBM package \cite{pysbm-github}, which generates SBM clusterings under a wide range of models. 
PySBM requires that the number of blocks be specified by the user.
We set this number using the number of blocks computed using graph-tool on the same network; see Supplementary Materials, Section B 
 for details.
\end{itemize}

Details of the software and commands used to generate SBM clustering for all approaches are provided in the Supplementary Materials, Section  B. 

\subsection*{Post-processing treatments to improve connectivity}

We post-processed clusters in two ways (below), each of which was designed to improve the edge-connectivity of the clusters in a given clustering. The software and commands used to post-process the clusters are detailed in Supplementary Materials, Section C.  
 
\paragraph{Connected Components (CC)}

Given a cluster that is internally disconnected, we replaced it with its connected components; thus, each single disconnected cluster is replaced by two or more smaller clusters that are connected.

\paragraph{Well-Connected Clusters (WCC)}

The WCC technique is a simplification of CM \cite{cm-journal} that does not re-cluster during the iterative process.  
Thus, given a clustering of a network and a threshold for well-connectedness (defined by the minimum edge-cut size), WCC checks each cluster to see if it is well-connected.
If so, it places the cluster in the output; else, it partitions the cluster into two smaller clusters based on its minimum edge cut. The process iterates until each cluster is well-connected. 

As with CM, we used the default threshold from \cite{cm-journal} for well-connectedness, which requires a cluster to have a minimum edge cut size that is strictly greater than the base-10 logarithm of the number of nodes in the cluster. 
 
To reduce the runtime, if the minimum edge-cut could be obtained by deleting a single edge that separates one node from the rest of the cluster, we used the partition produced by removing that edge. Otherwise, the minimum edge-cut was the balanced minimum edge-cut computed by the VieCut \cite{henzinger2018practical} software with the cactus data structure.

\subsection*{Evaluation}  

We evaluated clusterings produced by the SBMs with and without post-processing using CC and WCC  on both real-world and synthetic networks.

For a real-world network, we analyzed the connectivity of the estimated clusters, noting the proportion of clusters that are disconnected, poorly connected, and well-connected.  We also evaluated methods for SBM clustering under different models based on the description length they find (lower is better). The code for the computation of the description length is provided in Supplementary Materials, Section  D. 

For a synthetic network with ground-truth communities, we evaluated clustering accuracy using NMI, ARI, and AMI. 
We also computed clustering accuracy after restricting the network to a subset of the nodes based on membership in ground-truth clusters that had sufficient edge-density.
Specifically, for a density threshold $t$, we include a cluster $C$ if the number of edges in $C$ is greater than $t \binom{n}{2}$, where $n$ is the number of nodes in $C$.
We consider a singleton cluster to have density $0.0$. 
Finally, we also evaluate estimated clusterings using precision and recall, so that each ground-truth cluster or estimated cluster defines a set of pairs, and we compare the set of pairs for estimated and for ground-truth clusterings.
See Supplementary Materials, Section E 
for additional details.   

\subsection*{Infrastructure}
\label{sec:infrastructutre}

This work made use of the Illinois Campus Cluster, a computing resource operated by the Illinois Campus Cluster Program (ICCP) in conjunction with the National Center for Supercomputing Applications (NCSA).

Each method was allowed up to 72 hours of runtime, 256GB of RAM, and 16 cores of parallelism. If a clustering method failed to complete on the network, we reported this and provided the reason for the failure. 

\subsection*{Experiments}
We conducted five experiments:

\begin{itemize}
\item Experiment 1: We compared PySBM to graph-tool to determine the best approach for SBM-based clustering. Based on this experiment, we selected graph-tool for future study.
\item Experiment 2: We evaluated edge-connectivity of clusters produced by graph-tool SBM clusterings, both flat and nested, on real-world networks. 
\item Experiment 3: We evaluated the impact on clustering accuracy of our two treatments on SBM clusterings using graph-tool, both flat and nested, on synthetic non-bipartite networks with ground-truth community structure.
\item Experiment 4: We analyzed the influence of different components in the description length formula for DC-Flat in graph-tool.
\item Experiment 5: We evaluated the computational performance of our treatments on large real-world networks.
\end{itemize}

\section*{Results}\label{sec:results} 

\subsection*{Experiment 1: Comparing PySBM to graph-tool}

This experiment compared PySBM \cite{pysbm} to graph-tool on real-world networks with respect to the proportion of disconnected clusters and the ability to find clusterings with the minimum description length for a specific SBM model. 
The experiment was conducted on the $10$ smallest real-world networks, with the number of nodes between $906$ and $2115$ (see the list in Supplementary Materials, Section 
F. 

PySBM implements three inference algorithms for minimizing description length: Kernighan-Lin (KL-EM) \cite{kernighan70heuristic}, Metropolis-Hastings with 250,000 iterations (MHA-250k), and Peixoto's Agglomerative Heuristic (PAH) \cite{peixoto2014efficient}. Note that PAH is similar to the main inference technique implemented in graph-tool.

\begin{figure}[!h]
\includegraphics[width=\textwidth]{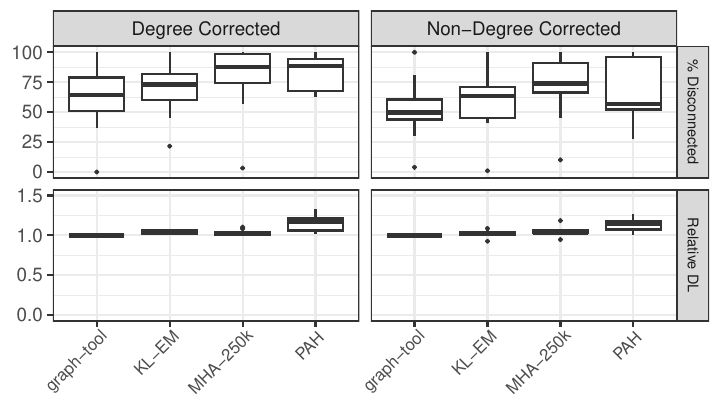}
\caption{\textbf{Experiment 1: Comparing PySBM and graph-tool} We compared three algorithmic techniques implemented in PySBM (KL-EM, MHA-250k, PAH) and graph-tool on the ten smallest real-world networks (between $906$ and $2115$ nodes in each network). 
The PySBM models use the number of blocks from graph-tool. 
Left: results for degree-corrected; right: results for non-degree-corrected;  top: percent of clusters that are disconnected; bottom: relative description length (DL) values compared to those found by graph-tool (values greater than $1.0$ indicate worse results compared to graph-tool).   
}
\label{fig:expt1}
\end{figure}

Fig~\ref{fig:expt1} shows the results for two SBM models that are implemented in both graph-tool and PySBM: degree-corrected (DC-Flat in graph-tool, DCPUH in PySBM) and non-degree-corrected (NDC-Flat in graph-tool, SPC in PySBM). 
We compared graph-tool's algorithm with the three PySBM's algorithms. 
The top row shows the percentage of clusters that are internally disconnected in the computed clustering. 
For the bottom row, we treated the description length of the clustering computed by the graph-tool's algorithm as the baseline and report the relative description length score, defined as the ratio of the description length of the clustering obtained by each inference algorithm to the baseline. 
Hence, the score for graph-tool's algorithm is set to $1.0$, and the score for PySBM's inference algorithms could be larger or smaller than $1.0$: values less than $1.0$ indicate that PySBM found a model with a better (lower) description length, and values greater than $1.0$ indicate that PySBM found a model with worse (higher) description length.
 
For both models studied (DC and NDC), although graph-tool and the best of PySBM's algorithms (KL and MHA-250k) had close description length scores, there was a slight advantage to graph-tool.
PAH had worse performance for description length.
In addition, for both models, graph-tool clusterings had a smaller percentage of disconnected clusters than those found by any of PySBM's techniques.

PySBM also implements additional SBM models beyond the two in common with graph-tool.  
In Supplementary Materials, Fig A 
we see that all of these models produced a high frequency of disconnected clusters.

Based on these observations, we restricted the rest of the study to using graph-tool.

\subsection*{Experiment 2: Evaluating edge-connectivity of graph-tool clusterings on real-world networks}

Table \ref{table:percent-disconn-realworld} shows the percentage of clusters that were internally disconnected in clusterings produced by five of graph-tool's SBM models on the $108$ small to moderate-sized real-world networks (see Materials and Methods).

On both bipartite and non-bipartite networks, a high percentage of the clusters were internally disconnected. Moreover, the percentage was much higher for bipartite networks than for non-bipartite networks.
Based on these observations, we restricted the rest of the study to the networks that are not bipartite.

On the non-bipartite networks,  all models produced at least 62.5\% disconnected clusters, with the highest percentage produced by DC-Flat and the lowest by PP-flat.
An in-depth examination of the connectivity of these clusterings is provided in
Supplementary,  
Figs B and C, 
which suggests that the incidence of disconnected clusters may increase as the network size increases, and that some models (e.g., PP-Flat) may tend to have more poorly-connected clusters than other models. 

\begin{table}[!ht]
\centering
\caption[Percentage of clusters that are disconnected for SBM models on real-world networks.]{\textbf{Experiment 2: Percentage of clusters that are disconnected for SBM models on real-world networks}}
\label{tab:disconn_perc}
\begin{tabular}{lrrrr}
\toprule
& \multicolumn{2}{c}{\textbf{Bipartite (34)}} & \multicolumn{2}{c}{\textbf{Non-Bipartite (74)}} \\
\cmidrule(lr){2-3}\cmidrule(lr){4-5}
Model & Mean & Std. Dev. & Mean & Std. Dev. \\
\midrule
NDC-Flat & 100.0 & 0.0 & 68.8 & 23.1 \\
DC-Flat & 100.0 & 0.0 & 76.1 & 23.8 \\
PP-Flat & 86.9 & 22.8 & 62.5 & 30.8 \\
NDC-Nested & 100.0 & 0.0 & 67.6 & 22.5 \\
DC-Nested & 100.0 & 0.0 & 72.4 & 24.6 \\
\bottomrule

\end{tabular}
\begin{flushleft}
We show the mean and standard deviation of the percentage of internally disconnected clusters across $34$ bipartite networks (left) and $74$ non-bipartite networks (right) for $5$ different SBM models implemented in graph-tool.
\end{flushleft}
\label{table:percent-disconn-realworld}
\end{table}

\subsection*{Experiment 3: Impact of treatment on synthetic networks}

Experiment 3 comprises two separate experiments.
In Experiment 3a, we  compared the clustering accuracy of Chosen-Nested SBM to Chosen-Flat SBMs, exploring the impact of treatment using WCC.
This experiment includes all four types of EC-SBM networks, which vary based on the input clustering (SBM+CC, Leiden-CPM(0.1), Leiden-CPM(0.01), and Leiden-Mod). 

In Experiment 3b,  we examined the impact of both CC and WCC treatments on Chosen-Nested and Chosen-Flat SBM clustering.
Results shown here for Experiment 3b are only for Chosen-Nested and  EC-SBM networks based on input clusterings Leiden-CPM(0.1) and Leiden-Mod; results for all EC-SBM networks are 
 shown in the Supplementary Materials, Figs D--H. 

\subsubsection*{Experiment 3a: Comparing Chosen-Flat vs.~Chosen-Nested}
 
\begin{figure}[!htpb]
    \centering
    \includegraphics[width=0.9\textwidth]{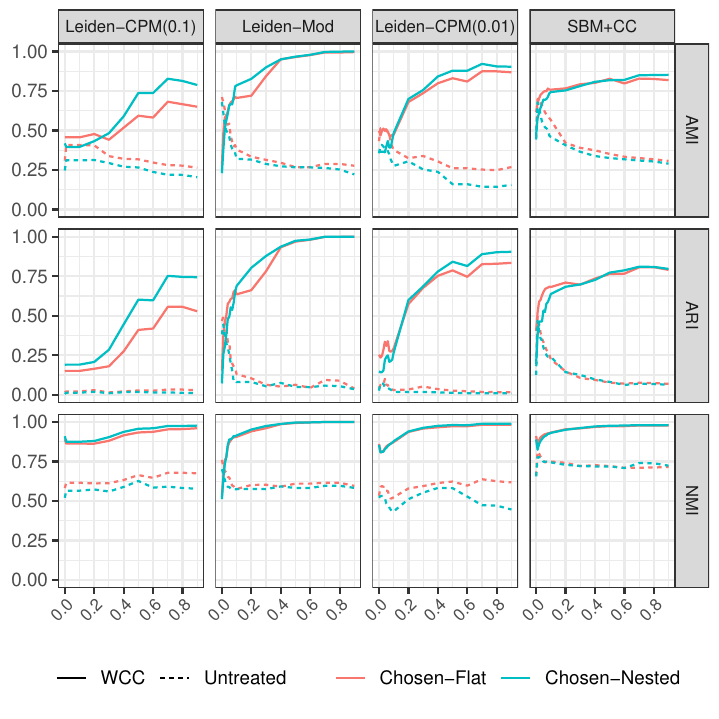}
    \caption[Experiment 3: Comparison between Chosen-Flat and Chosen-Nested on community detection accuracy]{\textbf{Experiment 3: Comparison between Chosen-Flat and Chosen-Nested on community detection accuracy}  
    Each column corresponds to a different set of EC-SBM networks, distinguished by the input clustering method specified in the column label. Specifically, from left to right, there are $73, 71, 73,$ and $74$ networks (missing \texttt{myspace\_aminer} for Leiden-CPM at both resolutions and additionally \texttt{berkstan\_web} and \texttt{petster} for Leiden-Mod, due to memory issues for WCC on these networks).
    Each row corresponds to an accuracy metric, and the value shown for each threshold is the median across all networks. The $x$-axis indicates a density threshold. For each threshold value, only clusters with a density strictly greater than the threshold value are analyzed. 
    Here, singleton clusters are included in the $0.0$ threshold.
    The results show that WCC improves accuracy for both Chosen-Flat and Chosen-Nested in nearly all cases, with only a few exceptions when ground-truth clusters with the lowest density are included. They also show that untreated Chosen-Flat is generally more accurate than untreated Chosen-Nested, but Chosen-Nested+WCC is more accurate than Chosen-Flat+WCC.
    }
    \label{fig:expt3a}
\end{figure}

Experiment 3a evaluated the impact of WCC treatment on Chosen-Nested and Chosen-Flat SBM clusterings produced using graph-tool, for all four types of EC-SBM networks, as defined by the input clustering.
There are 12 subfigures within
Fig \ref{fig:expt3a}, specified by the accuracy criterion (row) and type of EC-SBM networks (column).  
The x-axis within each subfigure indicates the minimum required density for the clusters under consideration, so that the evaluation is with respect to a varying set of clusters: all the clusters (even singleton clusters) are considered when the minimum density is $0.0$, but the set decreases in size to remove the sparse clusters as $x$ increases.
Each subfigure has curves for clusterings based on Chosen-Flat and Chosen-Nested,  both WCC-treated and untreated.

The impact of WCC treatment on Chosen-Nested or Chosen-Flat depended very much on the EC-SBM network type, as defined by the input clustering. 
For Leiden-CPM(0.1) input clusterings, WCC treatment improved SBM clustering accuracy for both Chosen-Flat and Chosen-Nested for all three accuracy criteria, and the benefit increased with the density threshold.
For Leiden-CPM(0.01) input clusterings, WCC treatment improved SBM clustering accuracy for both Chosen-Flat and Chosen-Nested when considering either ARI or NMI, but had a variable impact on AMI accuracy.
Specifically, WCC improved AMI accuracy for Chosen-Flat both overall and at every density threshold, but reduced overall AMI accuracy for Chosen-Nested.
Interestingly, WCC also improved AMI accuracy for Chosen-Nested when restricted to the ground-truth clusters with density above about $0.1$.
Results on EC-SBM models based on SBM+CC and Leiden-Mod show different trends:  WCC treatment hurt overall accuracy for both Chosen-Flat and Chosen-Nested under all three accuracy criteria.
However, after restriction to ground-truth clusters with density above about $0.1$, WCC treatment improved accuracy for both Chosen-Flat and Chosen-Nested under all three accuracy criteria.

This suggests that the reduction in overall accuracy caused by WCC treatment was due to poor accuracy on the sparsest clusters (those with density below $0.1$), and that WCC treatment improved accuracy when clusters had higher density.
This question is addressed in Experiment 3b.

\subsubsection*{Experiment 3b: Examining CC and WCC on graph-tool SBMs}

The goal of Experiment 3b was to understand the impact of both CC and WCC on graph-tool SBM clustering. 
Experiment 3a provided some insight into this question, but did not examine the impact of WCC, and also did not adequately explore the impact of either CC or WCC on overall accuracy.  Hence, in this experiment, we provide a more in-depth analysis.

\paragraph{Impact of CC and WCC treatment}

While results for all examined graph-tool SBM models are provided in the Supplementary Materials, we present Chosen-Nested here, since it had the best accuracy in Experiment 3a. 
We consider two types of EC-SBM networks: those based on Leiden-CPM(0.1) and those based on Leiden-Mod input clusterings, as they represent two extremes: the EC-SBM models based on Leiden-CPM(0.1) are the ones that had the best fit to the real-world networks, and the EC-SBM models based on Leiden-Mod have the worst fit to the real-world networks \cite{vu2025ecsbm}.  

\begin{figure}[!htpb]
    \centering
    \includegraphics[width=0.9\textwidth]{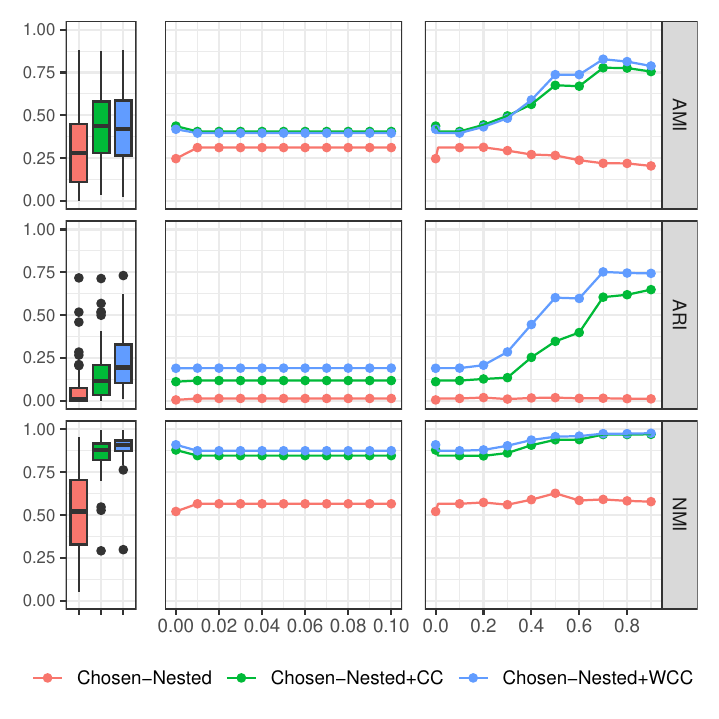}
    \caption[Experiment 3: Effect of treatments on accuracy of Chosen-Nested on EC-SBM Leiden-CPM(0.1) networks]{\textbf{Experiment 3: Effect of treatments on accuracy of Chosen-Nested on EC-SBM Leiden-CPM(0.1) networks} The results are from $73$ EC-SBM networks using Leiden-CPM(0.1) for input clustering (missing \texttt{myspace\_animer} due to memory issue for WCC on these networks). Each row corresponds to an accuracy metric, and values shown are medians across all networks. 
    The leftmost column shows the accuracy when all clusters in each network are considered. For the middle and rightmost columns, the $x$-axis represents a density threshold: in these columns, only clusters with a density strictly greater than the threshold are analyzed. 
    Here, singleton clusters are included in the $0.0$ threshold.
    Both CC and WCC treatments improve accuracy for all criteria for Chosen-Nested, and the degree of improvement increases with the density of the cluster.
    }
    \label{fig:expt3b-leidencpm}
\end{figure}

We begin with results on the EC-SBM networks based on Leiden-CPM(0.1) clusterings (Fig \ref{fig:expt3b-leidencpm}).
For all accuracy criteria, both CC and WCC treatments improved accuracy for Chosen-Nested, and the degree of improvement increased with the density of the cluster. 
However, WCC provided a larger improvement in accuracy compared to CC  
(most evidently for ARI accuracy). 
The same trend is observed on EC-SBM networks based on Leiden-CPM(0.01) and SBM+CC clusterings (see 
Supplementary Materials, Figs  D and E), 
as well as when using the Chosen-Flat model (see Supplementary Materials, Figs F--H. 

\begin{figure}[!htpb]
    \centering
    \includegraphics[width=\textwidth]{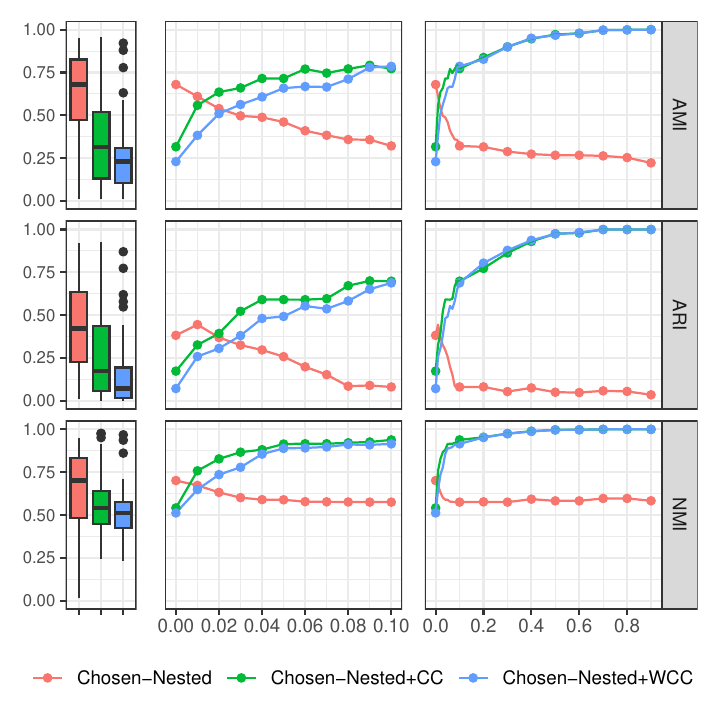}
    \caption[Effect of treatments on accuracy of Chosen-Nested on EC-SBM Leiden-Mod networks]{\textbf{Experiment 3: Effect of treatments on accuracy of Chosen-Nested on EC-SBM Leiden-Mod networks} The results are from $71$ EC-SBM networks using Leiden-Mod for input clustering (missing \texttt{myspace\_aminer}, \texttt{berkstan\_web}, and \texttt{petster}, due to memory issues for WCC on these networks).  
    Each row corresponds to an accuracy criterion, and values shown are medians across all networks.  The leftmost column shows the accuracy when all clusters in each network are considered. For the middle and rightmost columns, the $x$-axis represents a density threshold. In these columns, only clusters with a density strictly greater than the threshold are analyzed. 
    Here, singleton clusters are included in the $0.0$ threshold.
    Both CC and WCC treatments generally improve accuracy for Chosen-Nested for clusters with density at least $0.03$, but can reduce accuracy for the sparsest clusters (density at most $0.02$).
    }
    \label{fig:expt3b-leidenmod}
\end{figure}

Results on EC-SBM networks with Leiden-Mod input clusterings show somewhat different trends (Fig \ref{fig:expt3b-leidenmod}).
Both CC and WCC treatments improved accuracy for Chosen-Nested for clusters with density at least $0.03$, but sometimes reduced accuracy for the sparsest clusters (density at most $0.02$).
On the highest-density clusters (those with density at least $0.4$), CC and WCC treatments were equal in terms of accuracy. The same trend is observed when using the Chosen-Flat model (see Supplementary Materials, Fig I). 

An examination of the impact of WCC on overall clustering accuracy, which considers all clusters, shows somewhat different trends.
As noted in Fig \ref{fig:expt3b-leidencpm} (left-most column), WCC treatment improved overall accuracy under all three criteria for Chosen-Nested SBM clusterings of EC-SBM networks based on Leiden-CPM(0.1).
In contrast, Fig \ref{fig:expt3b-leidenmod} (left-most column) shows WCC treatment reduced overall accuracy under all three criteria for Chosen-Nested SBM clusterings of EC-SBM networks based on
 Leiden-Mod.
 The impact of WCC on flat or nested SBM clusterings on the two other types of EC-SBM networks ranged from relatively neutral to beneficial, depending on the criterion and network type. 
Thus, WCC was generally neutral to beneficial for all SBM clusterings of EC-SBM networks,  except for those EC-SBM networks based on Leiden-Mod clusterings.

\paragraph{Understanding why WCC has variable impact}

\begin{figure}[!htpb]
    \centering
    \includegraphics[width=\textwidth]{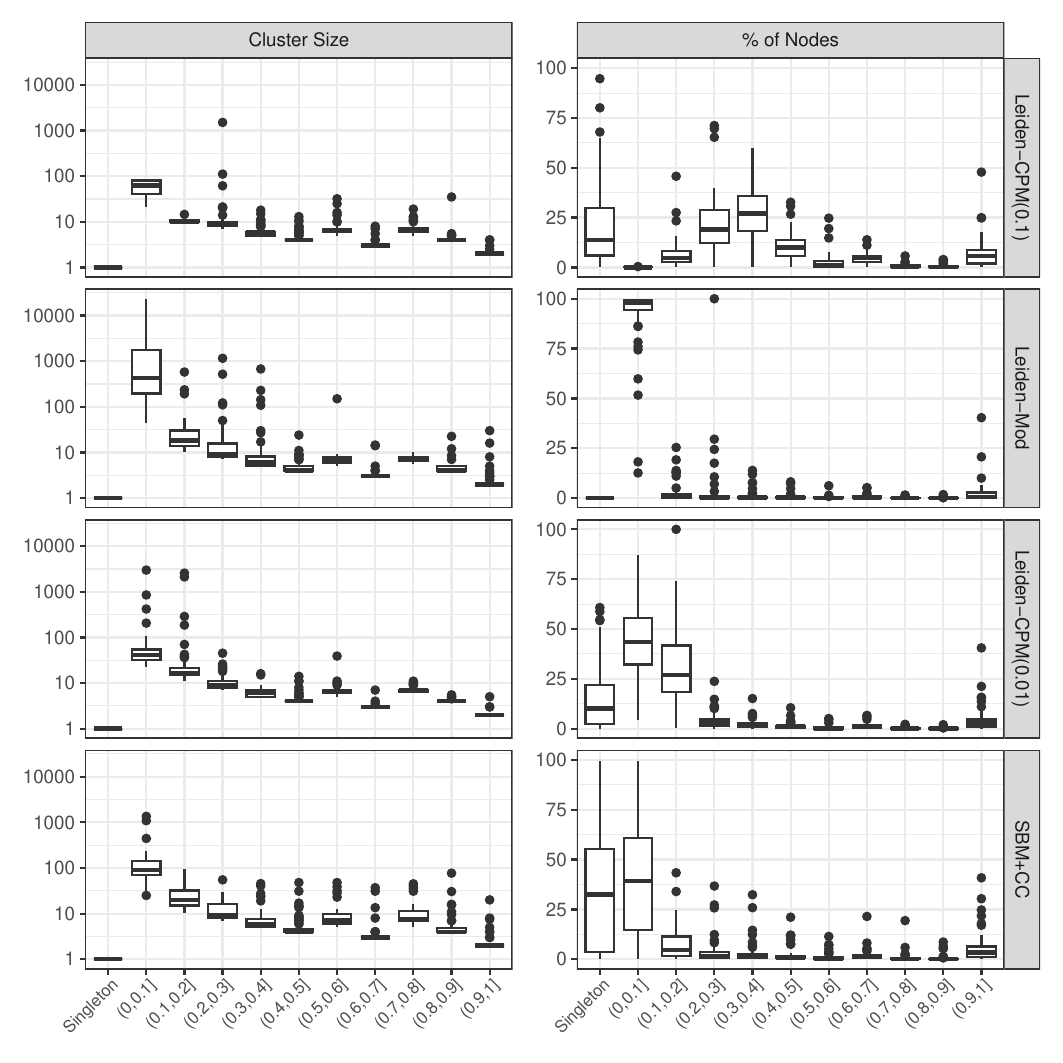}
    \caption[Statistics of EC-SBM ground-truth clusterings]{\textbf{Experiment 3: Statistics of EC-SBM ground-truth clusterings} Each row corresponds to a set of $74$ EC-SBM networks that are defined by the input clustering method given to EC-SBM, specified in the row label. 
    The clusters of each network are binned into sets: singleton clusters on the far left, followed by bins based on edge-density (i.e., proportion of edges present in the cluster out of the set of possible edges), from the least dense (i.e., edge-density at most 10\%) to the densest.
    The left column plots the median cluster size for each clustering, and the right column plots the percentage of nodes in that bin. 
    Supplementary Materials Figs J and K 
    provide results on finer-grained thresholding values.
    }
    \label{fig:cluster-size-median-density}
\end{figure}

To understand why the impact of WCC  depends on the type of EC-SBM network,  we investigated the properties of the ground-truth clusterings of the EC-SBM networks.
For each of the four types of EC-SBM networks, we computed the size and edge-density   (i.e., fraction of the edges that are present in each cluster) for the clusters.
After binning the clusters by edge-density, we observed features that are common across the different types of EC-SBM networks as well as striking differences  (see Fig \ref{fig:cluster-size-median-density} and finer-grained results in Supplementary Materials, Figs  J and K. 
For all EC-SBM network types, the cluster sizes generally decreased as the edge-density increased, and the highest-density clusters tended to be small.  
Beyond this, the EC-SBM networks differed in interesting ways. 
For example, with the exception of EC-SBM networks based on Leiden-Mod, which had very few singleton clusters (i.e., the median percentage of nodes in singleton clusters was zero), all other EC-SBM networks had singleton clusters, with median percentages ranging from about 10\% to about 30\%.

The EC-SBM networks also differed substantially in their density profiles. The EC-SBM networks based on Leiden-Mod clusterings had the vast majority (median value close to 100\%) of the nodes in the lowest density cluster bin, and these lowest density clusters were very large.
EC-SBM clusterings based on SBM+CC or Leiden-CPM(0.01) also had a reasonably large percentage of nodes in the lowest density bin, but the median value was much lower: between 40\% and 45\%. 
In significant contrast, EC-SBM networks based on Leiden-CPM(0.1) clusterings had very few clusters in the lowest density bin, and the bin with the largest number of nodes was relatively dense (i.e., edge-density between $0.3$ and $0.4$). 
The EC-SBM networks based on Leiden-CPM(0.1) clusterings also had the largest median fraction of nodes in the highest density bin. 


\begin{figure}[!ht]
    \centering
    \includegraphics[width=0.95\linewidth]{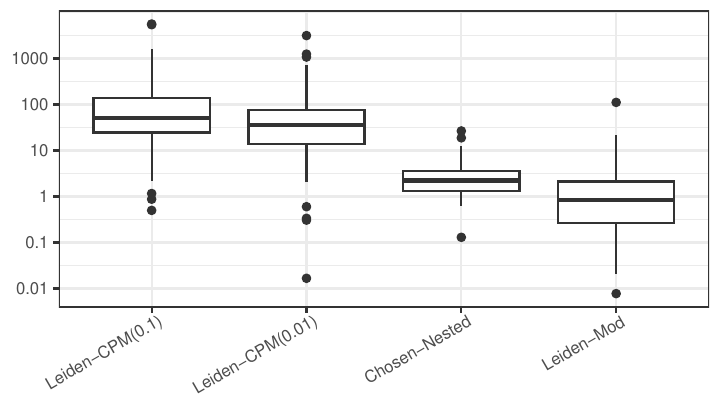}
    \caption[Normalized number of clusters produced by clustering methods on real-world networks]{\textbf{Experiment 3: Normalized number of clusters produced by clustering methods on real-world networks} For each of the $74$ real-world networks, we plot the number of clusters produced by each clustering method, normalized by that of Chosen-Flat. Leiden-Mod and Chosen-Flat produce similar numbers of clusters. Chosen-Nested, Leiden-CPM(0.01), and Leiden-CPM(0.1) produce many more clusters than Chosen-Flat. From left to right, the medians are at 51.1, 35.6, 2.23, and 0.831.
    Note that these values also reflect the number of ground-truth communities in the EC-SBM networks. Thus, EC-SBM networks based on Leiden-Mod have approximately the same number of ground-truth communities as EC-SBM networks based on Chosen-Flat.  EC-SBM networks based on Leiden-CPM(0.01) have many more communities, and EC-SBM networks based on Leiden-CPM(0.1) have the most communities.
    }
    \label{fig:emp_numclusters}
\end{figure}

Finally, the EC-SBM networks also differed in the number of ground-truth communities: as seen in Fig \ref{fig:emp_numclusters}, EC-SBM networks based on Leiden-Mod and Chosen-Flat have approximately the same number of ground-truth communities. On the other hand, EC-SBM networks based on Leiden-CPM(0.01) have approximately 35.6 times as many ground-truth communities as EC-SBM networks based on Chosen-Flat, and EC-SBM networks based on Leiden-CPM(0.1) have roughly 51.1 times as many ground-truth communities as EC-SBM networks based on Chosen-Flat. 

Thus, the EC-SBM networks based on different input clusterings differ substantially in terms of the number of ground-truth communities and the density and size of these clusters.
While it is not clear exactly how each of these differences contributes to the impact of WCC, one obvious explanation is simply this. 
When the number of ground-truth communities is much larger than the number of communities that the SBM is able to detect (which may be much smaller due to the resolution limit), SBM will be forced to produce a smaller number of communities than the true number, and each of which will be less dense than the true communities. 
Using WCC on these clusters will tend to break them up into smaller and denser communities, thus increasing the number of communities and potentially improving the clustering accuracy.  
In contrast, when the number of true communities is close to the number that the SBM is able to return, then SBM clustering may produce a relatively accurate clustering. 
If that clustering contains poorly-connected clusters, then applying WCC will break up those clusters, thus potentially reducing accuracy. 
This explanation fits with the observation in this experiment where WCC was overall detrimental for EC-SBM networks based on Leiden-Mod, and otherwise was either beneficial or neutral.

\paragraph{Understanding why untreated SBM clustering has reduced accuracy as cluster density increases}

\begin{figure}
   \centering
    \includegraphics[width=\textwidth]{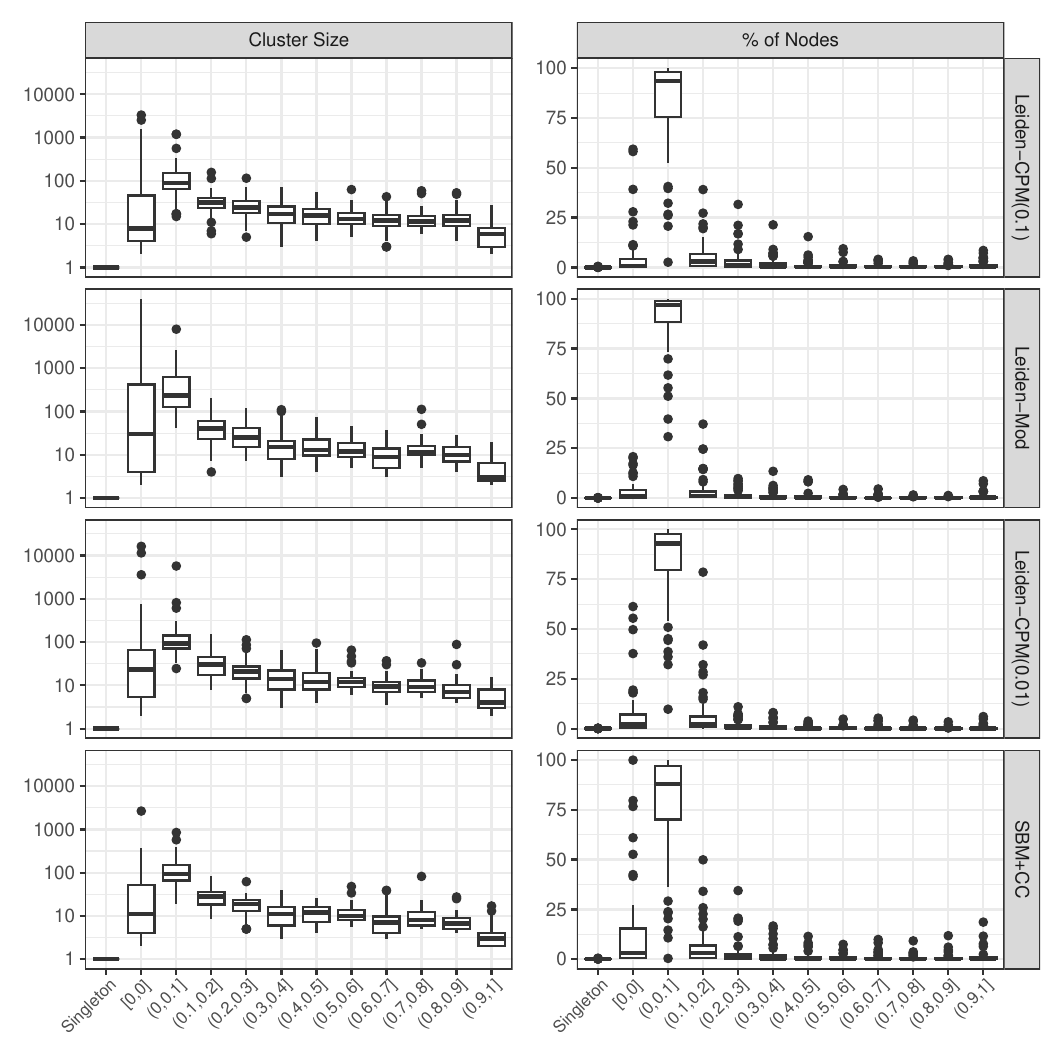}
   \caption{\textbf{Experiment 3: Statistics for Chosen-Nested clusterings on synthetic networks.}
   Each row indicates a collection of 74 EC-SBM networks that differ by the input clustering; the left column shows cluster sizes, and the right column shows the percentage of nodes, binned by cluster density range. For all synthetic networks, there are no singleton clusters, but many clusters have no edges (indicated by density $0.0$), and the vast majority of the nodes are in very sparse clusters (density at most $0.1$). Also, note that there is a small range of cluster sizes, but also some very small clusters (below size 10). 
   Supplementary Materials Fig L 
   provides these results for Chosen-Flat, which have similar trends.
   }
   \label{fig:stat-chosen-nested-synthetic}
\end{figure}

We comment on the observation that accuracy for untreated SBM clusterings, both flat and nested, often started high and then decreased as the density threshold increased (see Fig \ref{fig:expt3a}).  
An explanation for this may be seen by noting that ground-truth cluster size tended to decrease with the cluster density (Fig \ref{fig:cluster-size-median-density}), but Chosen-Flat and Chosen-Nested SBM clusterings of these synthetic networks had very few nodes in small clusters
(see Fig \ref{fig:stat-chosen-nested-synthetic} for the trends on Chosen-Nested; results on Chosen-Flat are shown in Supplementary Materials, Fig  L 
and display nearly identical trends).
For additional insight, we show precision and recall for Chosen-Flat and Chosen-Nested, with and without WCC treatment, in Supplementary Materials, Fig M.
These results show that while recall improved for all methods as density increased, precision tended to decrease for untreated Chosen-Flat and untreated Chosen-Nested, indicating that (to some extent) nodes from different dense ground-truth clusters were being merged into larger clusters.
In contrast, precision tended to improve with increases in density when Chosen-Flat and Chosen-Nested are WCC-treated.  
This improvement in recall and precision for WCC-treated methods also provides an insight into why applying WCC postprocessing improved accuracy when restricted to dense clusters: WCC breaks up the large sparse clusters into smaller dense clusters that are closer to the ground-truth clusters.

\clearpage

\subsection*{Experiment 4: Understanding why degree-corrected models produce disconnected clusters} 

It has already been noted that SBM clusterings have a type of resolution limit \cite{peixoto2019bayesian}; specifically, it will not be able to detect a number of blocks greater than $B_{\max} \propto \sqrt{n}$ where $n$ is the number of nodes in the network.
\cite{peixoto2019bayesian} gave an example of a network with many cliques as components where SBM clusterings produced clusters that contain two or more cliques, and hence will be internally disconnected. 
The explanation given in \cite{peixoto2019bayesian} for this phenomenon was that the description length for that model grows quadratically in the number of blocks (i.e., clusters), thus penalizing a large number of clusters. 

In \cite{park2024improved-conf-proceedings}, we extended that observation and derivation to the case of the DC-Flat SBM. 
We also examined description length scores for DC-Flat SBM clusterings on real-world networks and compared them to DC-Flat clusterings treated with CC; this analysis revealed that a specific component of the description length dominated the score when we use the CC post-processing treatment. For the sake of completeness, we provide this analysis here, along with our additional studies where we explore in greater depth the effect of the priors for the description length.

\paragraph{Analysis of the description length formula}

For a given input network $N$, graph-tool under the DC-Flat model seeks a clustering that minimizes the description length, making it an optimization problem. 

We now define the description length. Given a proposed SBM, let
\begin{itemize}
    \item $A$ be the adjacency matrix, defined by the network $N$,
    \item $b$ be the block (cluster) assignment, which represents the clustering of $N$,
    \item $k$ be the degree vector, defined by $A$,  
    \item $e$ be the edge count matrix, defined  by $A$ and $b$,
    \item $\beta \in [0, 1]$ be the weight of the priors (default to $\beta = 1.0$).
\end{itemize}
Eq~(\ref{eq:dcsbm-dl}) provides the formula for the description length $\mathrm{DL}(A, b)$ of a network $A$ and a clustering $b$ under the DC-Flat model:
\begin{eqnarray}
    \mathrm{DL}(A, b) = - \log p(A|b, e, k) - \beta \log p(k|b, e) - \beta \log p(b) - \beta \log p(e)
\label{eq:dcsbm-dl}
\end{eqnarray} 
For the following analysis, we consider the default configuration of $\beta = 1.0$. We will consider other configurations in a later analysis.

The description length can be decomposed into four parts, which are the negative logarithms of the model likelihood ($-\log p(A|b, e, k)$), the prior for the degree sequence ($-\log p(k|b, e)$), the prior for the block assignment ($-\log p(b)$), and the prior for the edge count matrix ($-\log p(e)$). graph-tool provides the functionality for computing these quantities separately, which we will use for the following analyses (see the software in Supplementary Materials, Section D). 

We compute the components of the description length for all networks that select the DC-Flat model, with and without the CC treatment; there are $65$ networks (see the full list in Supplementary Materials, Section G.1). 
Fig~\ref{fig:8} illustrates the distribution of differences between DC-Flat+CC (i.e., the output of the DC-Flat model with CC treatment) and DC-Flat (i.e., the output of the DC-Flat model) for each component on all the networks. Since the difference is DC-Flat+CC - DC-Flat, a positive difference means we do not favor the CC treatment. On all studied networks, the $-\log p(A|b, e, k)$ and $-\log p(k|b, e)$ component prefers connected clusters returned by the CC treatment. In contrast, the $-\log p(b)$ and $-\log p(e)$ components penalize the CC treatment, with the $-\log p(e)$ component contributing to a larger difference.

\begin{figure}[!h]
    \centering
    \includegraphics[width=0.8\linewidth]{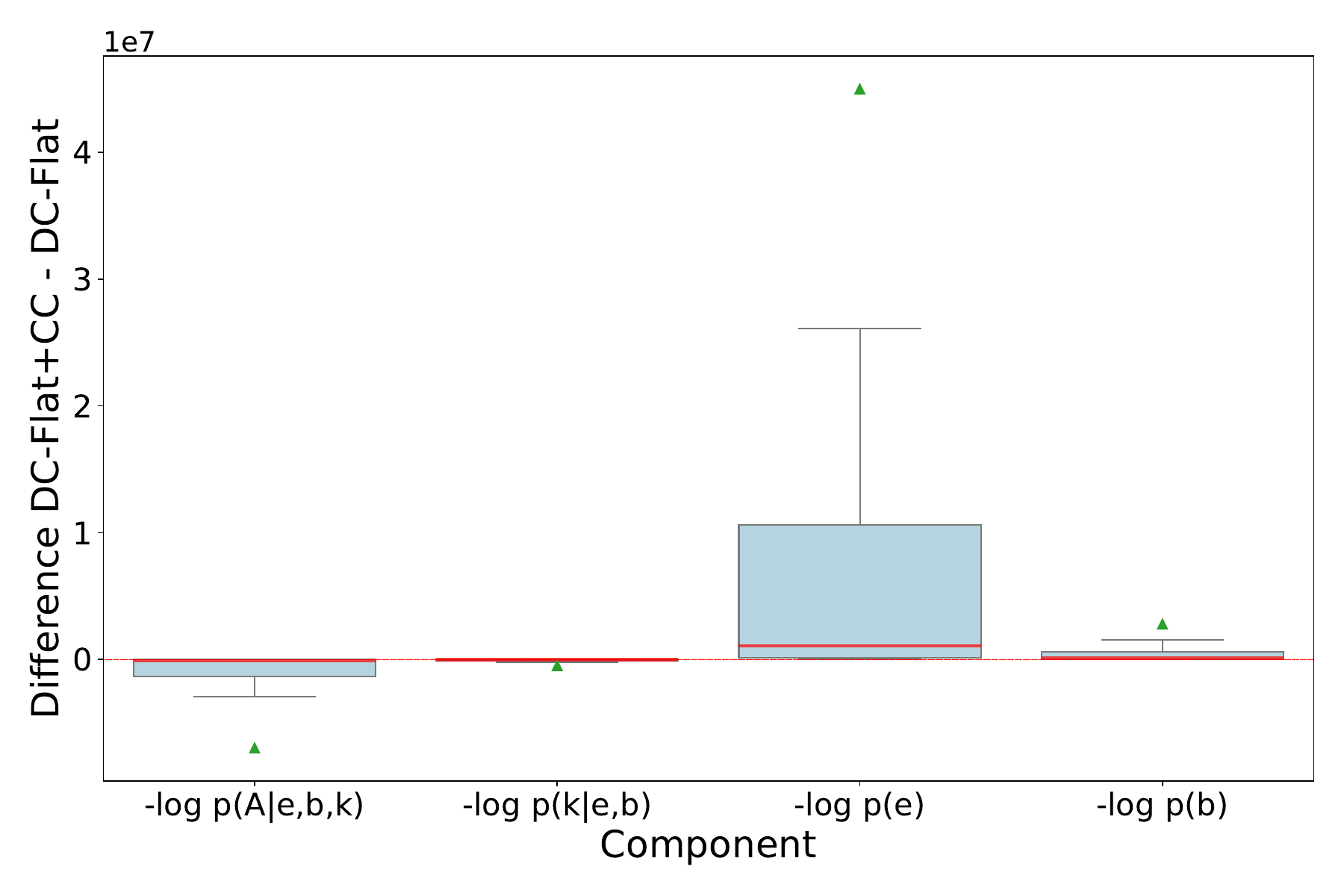}
    \caption[DC-Flat+CC to DC-Flat difference for description length components]{\textbf{Experiment 4: DC-Flat+CC to DC-Flat difference for description length components.} We show the contribution of $-\log p(e)$ term on the description lengths of CC-treated SBM clusterings. Box plots of differences for components of the description length on the $65$ networks that select the DC-Flat model for community detection and clusterings obtained by both DC-Flat and DC-Flat+CC for each network. The differences are DC-Flat+CC - DC-Flat. Positive values indicate favoring not having the CC treatment. 
    }
    \label{fig:8}
\end{figure}

We present a specific example of this phenomenon on the real-world network \texttt{linux}, detailed in Table \ref{tab:1}. The two components $-\log p(A|b, e, k)$ and $-\log p(k|b, e)$ of DC-Flat+CC are lower, indicating an advantage of the CC treatment. However, the two components $-\log p(b)$ and $-\log p(e)$ of DC-Flat+CC are higher, especially with the $- \log p(e)$ component having a much larger margin; this effectively negates all the advantage. As a result, with a lower description length, the untreated output of the DC-Flat model of SBM is preferred over the same output with CC treatment. Had the $- \log p(e)$ {\em not been included} in the formula for the description length, DC-Flat+CC would have been preferred for having a smaller description length.

\begin{table}[!ht]
\centering
\caption[Components of description lengths on \texttt{linux}.]{\textbf{Experiment 4: Components of description lengths on \texttt{linux}.}}
\label{tab:1}
\begin{tabular}{lrrr}
\hline
Quantity & SBM(DC) & SBM(DC)+CC & Difference \\
\hline
$-\log p(A|b, e, k)$ & 699k & 316k & -383k \\
$-\log p(k|b, e)$ & 96k & 45k & -51k  \\
$-\log p(b)$ & 147k & 257k & 110k \\
$-\log p(e)$ & 51k & 1,585k & 1,534k \\
\hline
DL$(A, b)$ & 993k & 2,202k & 1,209k \\
\hline
\end{tabular}
\begin{flushleft}
The last row sums up the values in the previous four rows. The difference is DC-Flat+CC - DC-Flat; so, a negative value means favoring DC-Flat with CC treatment, and a positive value means favoring untreated DC-Flat.
\end{flushleft}
\end{table}

We investigated all networks to see if the CC treatment is preferred when we remove the $-\log p(e)$ component. Our investigation shows, for $64$ networks out of $65$ networks (except \texttt{at\_migrations}), removing the $-\log p(e)$ component will result in a strictly lower description length for the clustering output with CC treatment. Even for \texttt{at\_migrations}, removing the $-\log p(e)$ component will result in the same description length for the clustering output with CC treatment. Thus, the $-\log p(e)$ component accounts for $100\%$ of the cases where DC-Flat without CC treatment is preferred over DC-Flat with CC treatment on the non-bipartite real-world networks we studied.

We now examine the formula for the $-\log p(e)$ component of DC-Flat. For undirected graphs, the formula is given by
\begin{eqnarray}
    - \log p(e) = \log \begin{pmatrix}
        B(B + 1)/2 + E - 1 \\
        E
    \end{pmatrix}
\end{eqnarray}
where $B$ is the number of blocks and $E$ is the number of edges. The formula shows that $-\log p(e)$ will increase as $B$ increases. 
Moreover, since $E>0$ is fixed,   $-\log p(e) = O(\log B)$.
Thus, this favors a small number of blocks $B$.

Recall that both the CC and WCC treatments tend to increase the number of blocks $B$, and that even CC increases this number whenever any of the clusters are disconnected.
Since increasing the number of blocks makes  $-\log p(e)$ larger, this means that such treatments will tend to result in larger description lengths---which will not be favored by SBM.
Since the description length is impacted strongly by $-\log p(e)$, this
explains why SBM clustering will tend to prefer clusterings that have internally disconnected clusters rather than their CC-treated versions.

Moreover, as seen in Fig \ref{fig:emp_numclusters}, although Leiden-Mod and Chosen-Flat produce roughly the same number of clusters on the 74 real-world networks,  Chosen-Nested produces more clusters, but  Leiden-CPM(0.1) and Leiden-CPM(0.01) produce far more clusters than even Chosen-Nested.
This supports the hypothesis that the pressure to produce a small number of clusters contributes to the appearance of internally disconnected clusters.

\paragraph{Studying the effect of the priors}
 
The previous discussion used the default configuration where $\beta = 1.0$ and found that certain influential prior components, especially $-\log p(e)$, encourage having large but poorly-connected clusters. In the following analysis, we experiment with different weight configurations in order to see if connectivity and clustering accuracy can be improved. 

The code for
graph-tool allows $\beta$ to be provided as an input parameter, and also allows each component to be ``turned off'' individually. 
For example, turning off the edge count matrix prior will lead to optimizing $\mathrm{DL}(A, b) + \beta \log p(e)$ instead of $\mathrm{DL}(A, b)$. Hence, we experiment with varying $\beta$ and turning off the edge count matrix component. The codes used are given in Supplementary Materials, Section G.2. 

Fig~\ref{fig:sbm_weighted_prior_real_conn} shows the effect of different prior weight configurations on the proportion of nodes in different types of clusters detected from real-world networks. Without the $p(e)$ component, almost all clusters are singletons. When varying $\beta$ from $0.0$ to $1.0$, the proportion of singleton clusters decreases while the proportion of nodes in disconnected clusters increases. There is a good value for $\beta$ in between $0.0$ and $1.0$ if we want a large part of the network to be in connected clusters ($\beta = 0.5$) or well-connected clusters ($\beta = 0.3$ or $0.4$).

\begin{figure}[!h]
    \centering
    \includegraphics[width=0.95\linewidth]{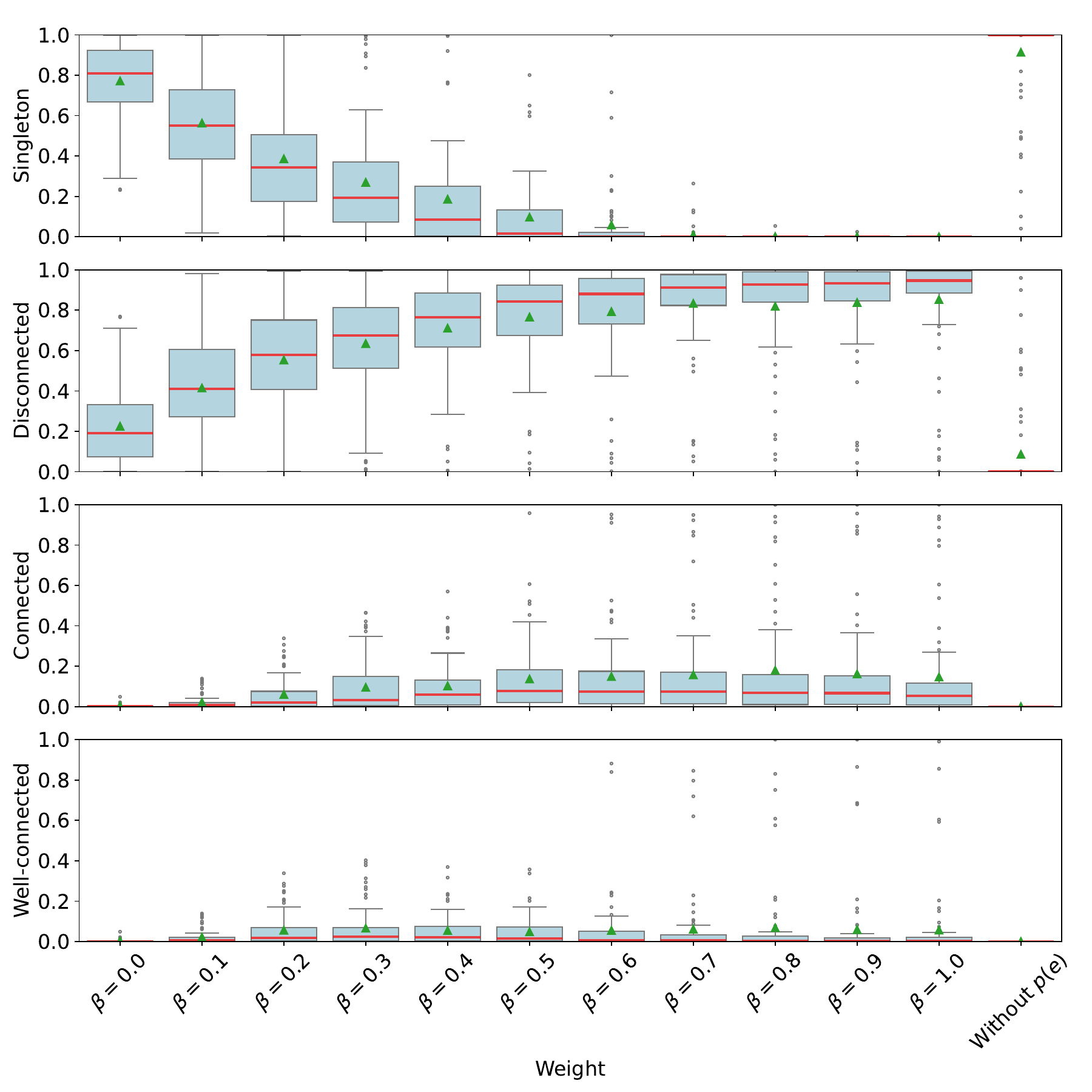}
    \caption[Effect of the prior weight configuration on the connectivity of detected clusters on real-world networks]{\textbf{Experiment 4: Effect of prior weight configuration on connectivity of detected clusters on real-world networks} The result is collected from $74$ real-world networks.  
    The $x$-axis shows different prior weight configurations. Each box shows the distribution across all networks of the proportions of nodes inside any cluster of the type indicated by the row label.}
    \label{fig:sbm_weighted_prior_real_conn}
\end{figure}

\begin{figure}[!ht]
    \centering
    \includegraphics[width=0.95\linewidth]{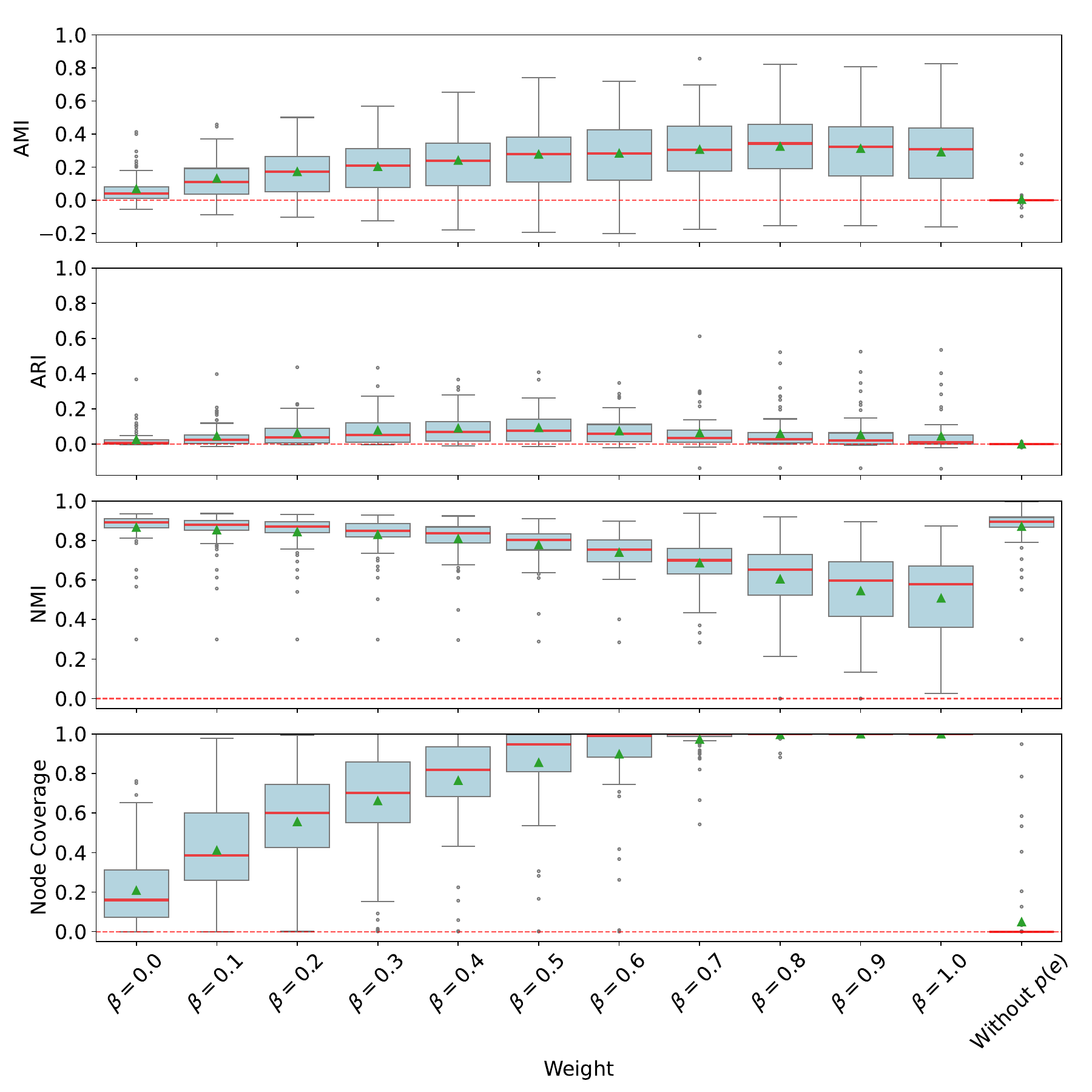}
    \caption[Effect of the prior weight configuration on community detection accuracy on synthetic networks]{\textbf{Experiment 4: Effect of prior weight configuration on community detection accuracy on synthetic networks} The result is collected from $74$ EC-SBM networks with Leiden-CPM(0.1) input clustering.  
     The $x$-axis shows different prior weight configurations. The values are accuracy metrics (top 3 rows) and proportion of nodes in singleton clusters (last row).}
    \label{fig:sbm_weighted_prior_syn_acc}
\end{figure}

Fig~\ref{fig:sbm_weighted_prior_syn_acc} shows the effect of prior weight configuration on the community detection accuracy for synthetic networks. Without the $-\log p(e)$ component (rightmost box), node coverage is $0\%$ for most networks, indicating that almost all clusters are singletons. When varying $\beta$ from $0.0$ to $1.0$ (boxes from left to right), NMI decreases while AMI increases to the peak at $\beta = 0.8$ before dropping slightly. A similar trend to AMI is observed for ARI, but with the peak at $\beta = 0.5$. The trend for NMI is partially explained by NMI's tendency to favor estimated clustering with more clusters, thus favoring clustering with a low node coverage (note that each unclustered node is considered a singleton cluster). On the other hand, the trends for AMI and ARI suggest that there are potential benefits for changing $\beta$ to a smaller value.

\clearpage

\subsection*{Experiment 5: Computational performance}

We examined the computational performance of the WCC treatment, specifically with respect to runtime and the memory required. The Chosen-Flat/Nested+WCC pipeline involves computing an SBM clustering under each model (DC-Flat, NDC-Flat, PP-Flat, DC-Nested, and NDC-Nested), and then following with WCC on the clustering that achieved the lowest description length. In all of our experiments, the SBM clustering step was completed, while WCC failed for some runs, as detailed below.

\subsubsection*{Performance on synthetic networks}
Within the given memory limit of 256GB and a time limit of 3 days, several runs on the synthetic networks did not complete. 
Note that all of these runs failed in the WCC treatment stage.
The Chosen-Flat model on Leiden-CPM networks with resolutions of $0.1$ and $0.01$ for the \verb+myspace_aminer+ network both ran out of memory. 
The Chosen-Flat model on Leiden-Mod networks for the \verb+petster+ network exceeded the time limit. 
Furthermore, several runs using the Chosen-Nested configuration on Leiden-Mod networks encountered segmentation faults (SEGFAULT), specifically those on the \verb+berkstan_web+, \verb+myspace_aminer+, and \verb+petster+ networks. 
A fix for the segmentation fault and an optimization for the memory usage are currently being undertaken. 
 
\subsubsection*{Performance on real-world networks}
\begin{figure}[!ht]
\includegraphics[]{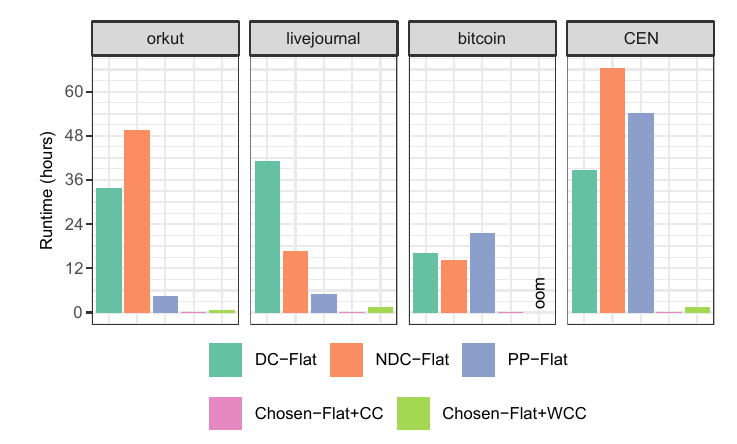}
\caption[Experiment 5: Runtime 
on large non-bipartite real-world networks]{\textbf{Experiment 5: Runtime of flat SBM with treatments on large non-bipartite real-world networks} 
For Chosen-Flat+CC and Chosen-Flat+WCC, only the time it took to run the treatment is shown. Chosen-Flat+WCC had an OOM on \texttt{bitcoin}. Number of nodes: \texttt{orkut} - 3,072,441; \texttt{livejournal} - 4,847,571; \texttt{bitcoin} - 6,336,770; \texttt{CEN} - 13,989,436.  The runtime of CC or WCC treatments on SBM clusterings is negligible compared to the runtime of SBM. 
} \label{fig:runtime}
\end{figure}

We show the runtime on the four largest real-world networks (see Fig~\ref{fig:runtime}). For this analysis, we limited each run to only a single core (i.e., no parallelization) for both graph-tool's inference and treatments. 
In general, the WCC treatment was able to complete on nearly every network we analyzed, except for one real-world network (\texttt{bitcoin}) due to an out-of-memory (OOM) error.

As shown in Fig~\ref{fig:runtime}, by far the most computationally intensive part was computing the SBM clustering for each of the three models, which took between 5 and 66 hours each. 
Running CC was negligible, completing in minutes, and running WCC finished in under two hours on each network. 
For example, on the CEN, which has about 14 million nodes, the SBM model with the least runtime was the DC-Flat model, which took 38.7 hours. 
In comparison, WCC processing took 1.4 hours. 
Other SBM models on the CEN were more expensive, with NDC-Flat at 66.5 hours and PP-Flat at 54.2 hours.
Thus, the time it took to cluster the networks using SBM far exceeded the time it took to process those clusterings through the CC and WCC treatments. 

\clearpage

\section*{Discussion} 
\label{sec:discussion}


A major finding from this study is that alternative models and software for SBM clustering did not succeed in reliably producing connected clusters. 
For example, we found that PySBM had the same problem with producing disconnected clusters as graph-tool when using the models that both tools enable.
Furthermore, although PySBM enables additional models compared to graph-tool, clusterings under these models also had a high frequency of disconnected clusters.
Thus, PySBM does not provide a solution to this problem.
Furthermore, although the improvement in description length was small, graph-tool was able to produce better (lower) description lengths than PySBM for models that both tools implemented. 
Together, these observations indicate that the main advantage of PySBM may be the additional models it enables, and that otherwise, graph-tool is a better choice than PySBM for clustering networks using SBMs.

Nevertheless, our study shows that all models within graph-tool produced disconnected clusters (Table \ref{table:percent-disconn-realworld}).
Flat models had a slightly higher frequency than nested (i.e., hierarchical models), and DC (degree-corrected) had a slightly higher frequency than NDC (non-degree-corrected). 
The lowest frequency was found by PP (planted partition), but its mean frequency of disconnected clusters is still 62\%. 
The observation that all models had a high frequency of disconnected clusters suggests that the  ``resolution limit'' exists for all these SBM models.

Our study evaluated the impact of post-processing using either CC or WCC, and found that both improve clustering accuracy on synthetic networks, especially with respect to the detection of dense ground-truth clusters. 
We also observed that untreated nested models had worse accuracy than untreated flat models, and that WCC-treated nested models had the best overall accuracy. 
Importantly, WCC-treated nested and flat models struggle with detecting sparse ground-truth clusters.
This difficulty resulted in reduced overall accuracy compared to untreated clusterings only when a very large part of the network is inside these sparse clusters, which, in our study, occurred only when Leiden-Mod was used as the input clustering to EC-SBM.  

Because SBM clusterings based on all models had a high frequency of disconnected clusters, we explored whether there are potential benefits in changing the weight configuration to reduce the influence of the prior components of the description length.
We reiterate the argument in \cite{park2024improved-conf-proceedings} that some components, especially the prior for the edge count matrix, heavily penalize having a large number of clusters, with the result that graph-tool's SBM models favor fewer rather than more clusters.
By reducing the influence of the prior by setting a lower weight for the prior components or removing the edge count matrix prior completely, we observed a tendency to make smaller clusters, to the extent of making each node a singleton cluster. This helps some accuracy metrics (NMI) but hurts others (AMI and ARI). 
However, there are weight values between $0.0$ and $1.0$ for the collection of networks that we analyze (i.e., EC-SBM synthetic networks with Leiden-CPM(0.1) input clustering), where overall improvement in clustering accuracy may be obtained; this trend should be investigated in future work.

Finally, our evaluation of the computational performance of CC and WCC treatments on four large networks with up to $\sim 14$ million nodes shows that they are very fast, running in minutes, and thus a fraction of the hours or days of analysis needed to compute the SBM clusterings.  
Nevertheless, the WCC software exhibited memory issues,  which we are currently addressing. 


This study extends the earlier conference paper  \cite{park2024improved-conf-proceedings} on community detection using stochastic block models. Since \cite{park2024improved-conf-proceedings} was limited to graph-tool SBM clusterings using flat models, whereas this study additionally examined nested models in graph-tool as well as clustering using PySBM software, this study provides additional insight into SBM clustering beyond what was shown in \cite{park2024improved-conf-proceedings}.

The most important observation in \cite{park2024improved-conf-proceedings} was that flat SBM models in graph-tool produce disconnected clusters. 
Because we explored nested models as well as PySBM models, which \cite{park2024improved-conf-proceedings} did not, this study shows that these alternative approaches for SBM clustering also tend to produce disconnected clusters; thus, our study extends that observation from \cite{park2024improved-conf-proceedings} to a larger class of SBM models and software. 
Thus, our study supports this important observation.

The second most important observation in \cite{park2024improved-conf-proceedings} is that WCC improves clustering accuracy. 
Our study is based on a larger set of synthetic networks (produced using EC-SBM \cite{vu2025ecsbm} instead of LFR \cite{lfr}), which allows us to provide a careful evaluation of this claim. To some extent, we support this observation. However, our study also shows that the impact of WCC depends on the properties of the ground-truth clusters in the synthetic network. 
Our study indicates that WCC helps when the number of ground-truth clusters is larger than what the SBM is able to recover, which indicates the presence of the ``resolution limit", but can hurt overall accuracy when the number of ground-truth communities is small enough (e.g., not more than what SBM is able to recover).
Thus, this study provides a nuanced conclusion regarding the impact of WCC on clustering accuracy and emphasizes the importance of the choice of synthetic networks when evaluating clustering methods and pipelines.  

Other observations in this study compared to \cite{park2024improved-conf-proceedings} are that graph-tool's nested models, followed by WCC, provide the best overall accuracy of all tested methods, when the synthetic network is not characterized by having nearly all the nodes in very large sparse ground-truth clusters.   
We also found that WCC greatly improved recovery of dense clusters, with the improvement increasing with the density threshold. 
Finally, unlike \cite{park2024improved-conf-proceedings}, we experimented with different weight configurations for the DC-Flat models and observed that improvements in the connectivity of the clusters and clustering accuracy can be achieved for modified settings of the weight configurations. 

\section*{Conclusions} \label{sec:conclusion}

Building on the work of \cite{park2024improved-conf-proceedings}, this paper presents an extensive study of community detection using Stochastic Block Models (SBMs). We evaluated a wide range of SBM software and their associated inference algorithms, testing various models, either in their original implementations and with our own modifications, across a large corpus of synthetic networks.

Our study revealed that the issue of producing disconnected clusters is present for all tested SBM models, including the ten models explored in PySBM and the five models explored in graph-tool, along with their respective inference algorithms. Although nested SBMs have been hypothesized to mitigate this problem, we found that they do not provide an adequate solution.  

The CC (connected components) technique is a simple approach that directly addresses the problem where a clustering has internally disconnected clusters, while the WCC (well-connected clusters) technique addresses the problem of poorly-connected clusters. 
Each of these two techniques was demonstrated to improve clustering accuracy, especially for detecting dense clusters, {\em except} when the vast majority of nodes are in large and very sparse ground-truth clusters. 
The observed improvement implies that the original SBMs suffer from issues related to the resolution limit, causing them to group multiple dense subgraphs into a single large cluster that is less dense. 
Under these conditions, CC and WCC achieve better accuracy precisely because they decompose these large clusters. 

As noted above, the impact on accuracy depends on the ground-truth communities, so that when the network has a relatively small number of communities, the WCC treatment can be detrimental. 
On the other hand, using SBM without WCC when the true number of communities is larger can produce communities that are sparser than the true communities, and using WCC can fix this issue. 
Since in general it is not possible to know the correct number of communities, the decision of whether to use WCC for post-processing SBM clustering should depend on factors such as if the recovery of all communities is needed, or only the sufficiently dense communities. 
If only the sufficiently dense clusters are relevant to the research question, then WCC is helpful, as it improves recovery of these clusters. 
However, when all clusters, even those with very low density, are desired, then WCC may be beneficial or detrimental, depending on the community structure for the network: if most nodes are in reasonably dense clusters, then WCC should be generally beneficial, but if the majority of nodes are in very sparse clusters, it may be detrimental to use WCC.

Future research related to SBM clustering should explore modifications to the priors in the description length formulas to investigate whether changes can reduce the tendency to disconnected clusters while maintaining or improving clustering accuracy. One potential approach is a granular analysis on the effect of individual prior weights, potentially leading to an adaptive scheme where the properties of the input networks are taken into consideration in setting the weights.
Another approach is to redesign the priors to explicitly penalize disconnectivity and reward well-connectedness.
Each of these approaches would need to be explored both theoretically and empirically to provide reliable improvements in community detection.
Other future work should address the memory issues of the WCC software; as WCC depends on VieCut \cite{henzinger2018practical}, this issue may be solved through changes to how the minimum cuts are found.
In addition, future work should investigate the impact of WCC on other clustering methods. 
Finally, although the impact of WCC post-processing has been studied using both LFR and EC-SBM synthetic networks, its impact under other synthetic networks such as ABCD+o \cite{kaminski2023abcdo}, should also be examined.

\section*{Supporting information}
\textbf{Supplementary Materials Document}  This  PDF document contains additional details about the study.  

\section*{Funding} \label{sec:funding}
This work was supported in part by the Illinois-Insper partnership and the US National Science Foundation grant 2402559 (to TW and GC). The authors thank the Illinois Computes Program for allocations of cluster computing time.

%
%
%




\section*{Declarations}

\subsection*{Availability of data and materials} 

The real-world networks on which the analyses are based are already in the public domain. 
The EC-SBM networks are available at \cite{ecsbm-networks}. 
The software used is in the public domain.
The analysis scripts are available at \cite{the-anh-scripts}.
The commands used to perform analyses are provided in the Supplementary Materials.

\subsection*{Competing interests}
The authors declare that they have no competing interests.


\subsection*{Authors' contributions}

TVL provided the EC-SBM networks and evaluated graph-tool clustering methods using both real and EC-SBM synthetic networks, analyzed the data, and wrote the first draft of the manuscript.
MP developed the codes for CC and WCC and assisted in writing the first draft.
IC evaluated PySBM and SBMs within graph-tool using both real and EC-SBM synthetic networks, analyzed the data, and assisted in writing the first draft.
GC supervised the research, evaluated experimental results, and edited the drafts of the manuscript.
TW supervised the research, evaluated experimental results, and edited the drafts of the  manuscript.
\subsection*{Acknowledgments}
The authors thank the members of the Warnow-Chacko lab for helpful feedback.

\bibliography{clustering}
\end{document}


%
\title{Supplementary Materials for Stochastic Block Models for Community Detection}
%
%


\author[1]{\fnm{The-Anh} \sur{Vu-Le}}\email{vltanh@illinois.edu}
\equalcont{These authors contributed equally to this work.}
\author[1]{\fnm{Minkhyuk} \sur{Park}}\email{minhyuk2@illinois.edu}
\equalcont{These authors contributed equally to this work.}

\author[1]{\fnm{Ian} \sur{Chen}}\email{ianchen3@illinois.edu}

\author[1]{\fnm{George} \sur{Chacko}}\email{chackoge@illinois.edu}

\author*[1]{\fnm{Tandy} \sur{Warnow}}\email{warnow@illinois.edu}

\affil[1]{\orgdiv{Siebel School of Computing and Data Science}, \orgname{University of Illinois Urbana-Champaign}, \orgaddress{\street{201 N. Goodwin Avenue}, \city{Urbana}, \postcode{61801}, \state{IL}, \country{USA}}}

\date{}


\maketitle 


\tableofcontents
\listoffigures

\clearpage

\section{Real-world Networks}
\label{apdx:real-nets}

\paragraph{Networks (110) from \cite{netzschleuder} sorted by increasing node count}

\noindent Bipartite networks (34): \texttt{plant\_pol\_robertson}; \texttt{escorts}; \texttt{movielens\_100k}; \texttt{nematode\_mammal}; \texttt{paris\_transportation}; \texttt{jester}; \texttt{dbpedia\_writer}; \texttt{digg\_votes}; \texttt{dbtropes\_feature}; \texttt{dbpedia\_starring}; \texttt{github}; \texttt{dbpedia\_recordlabel}; \texttt{dbpedia\_producer}; \texttt{dbpedia\_location}; \texttt{dbpedia\_occupation}; \texttt{dbpedia\_genre}; \texttt{discogs\_label}; \texttt{wiki\_article\_words}; \texttt{corporate\_directors}; \texttt{lkml\_thread}; \texttt{bookcrossing}; \texttt{flickr\_groups}; \texttt{visualizeus}; \texttt{dbpedia\_country}; \texttt{stackoverflow}; \texttt{eu\_procurements}; \texttt{epinions}; \texttt{citeulike}; \texttt{dbpedia\_team}; \texttt{bibsonomy}; \texttt{reuters}; \texttt{discogs\_affiliation}; \texttt{amazon\_ratings}; \texttt{dblp\_author\_paper} \\

\noindent Non-bipartite networks (76): \texttt{dnc}; \texttt{uni\_email}; \texttt{polblogs}; \texttt{faa\_routes}; \texttt{netscience}; \texttt{new\_zealand\_collab}; \texttt{collins\_yeast}; \texttt{interactome\_stelzl}; \texttt{bible\_nouns}; \texttt{at\_migrations}; \texttt{interactome\_figeys}; \texttt{us\_air\_traffic}; \texttt{drosophila\_flybi}; \texttt{fly\_larva}; \texttt{interactome\_vidal}; \texttt{openflights}; \texttt{bitcoin\_alpha}; \texttt{fediverse}; \texttt{power}; \texttt{advogato}; \texttt{bitcoin\_trust}; \texttt{jung}; \texttt{reactome}; \texttt{jdk}; \texttt{elec}; \texttt{chess}; \texttt{sp\_infectious}; \texttt{wiki\_rfa}; \texttt{dblp\_cite}; \texttt{anybeat}; \texttt{chicago\_road}; \texttt{foldoc}; \texttt{inploid}; \texttt{google}; \texttt{marvel\_universe}; \texttt{fly\_hemibrain}; \texttt{internet\_as}; \texttt{word\_assoc}; \texttt{cora}; \texttt{lkml\_reply}; \texttt{linux}; \texttt{topology}; \texttt{email\_enron}; \texttt{pgp\_strong}; \texttt{facebook\_wall}; \texttt{slashdot\_threads}; \texttt{python\_dependency}; \texttt{marker\_cafe}; \texttt{epinions\_trust}; \texttt{slashdot\_zoo}; \texttt{twitter\_15m}; \texttt{prosper}; \texttt{wiki\_link\_dyn}; \texttt{livemocha}; \texttt{wikiconflict}; \texttt{lastfm\_aminer}; \texttt{wiki\_users}; \texttt{wordnet}; \texttt{douban}; \texttt{academia\_edu}; \texttt{google\_plus}; \texttt{libimseti}; \texttt{email\_eu}; \texttt{stanford\_web}; \texttt{dblp\_coauthor\_snap}; \texttt{notre\_dame\_web}; \texttt{citeseer}; \texttt{twitter}; \texttt{petster}; \texttt{yahoo\_ads}; \texttt{berkstan\_web}; \texttt{myspace\_aminer}; \texttt{google\_web}; \texttt{ hyves}; \texttt{livejournal}; \texttt{bitcoin}

\paragraph{Other networks (2) from \cite{cm-journal}}

\noindent{}\textit{non-bipartite}: \texttt{orkut}; \texttt{CEN} (Curated Exosome Network) \\

\clearpage

\section{Software and Commands for Clustering}
\label{suppl:software-commands/cluster}

\paragraph{Flat and Nested SBM}

\verb+run_flat_sbm.py+ and \verb+run_nested_sbm.py+ is available at \url{https://github.com/illinois-or-research-analytics/network-analysis-code}, commit ac9af43.

To compute the flat SBM clusterings, the command is
\begin{lstlisting}
python run_flat_sbm.py \
    --input-network <network> \
    --inner-sbm-model <model> \
    --output-prefix <output_prefix> \
    --output-clustering <output_clustering> \
    --num-processors <num_processors>
\end{lstlisting}
where 
\begin{itemize}
    \item \texttt{<network>} is the path to the edge list file of the network we want to cluster,
    \item \texttt{<model>} is the flat model we want to use (DC-SBM for DC-Flat, Non-DC-SBM for NDC-Flat, and PP-SBM for PP-Flat),
    \item \texttt{<output\_prefix>} is the directory to store intermediate files
    \item \texttt{<output\_clustering>} is the file to store the final clustering
    \item \texttt{<num\_processors>} is the number of CPUs to be used
\end{itemize}

To compute the nested SBM clusterings, the command is
\begin{lstlisting}
python run_nested_sbm.py \
    --input-network <network> \
    --inner-sbm-model <model> \
    --output-prefix <output_prefix> \
    --output-clustering <output_clustering> \
    --num-processors <num_processors>
\end{lstlisting}
where 
\begin{itemize}
    \item \texttt{<network>} is the path to the edge list file of the network we want to cluster,
    \item \texttt{<model>} is the flat model we want to use (DC-SBM for DC-Nested and Non-DC-SBM for NDC-Nested),
   \item \texttt{<output\_prefix>} is the directory to store intermediate files
    \item \texttt{<output\_clustering>} is the file to store the final clustering
    \item \texttt{<num\_processors>} is the number of CPUs to be used
\end{itemize}

\clearpage

\paragraph{PySBM}

PySBM~\cite{pysbm} is available at \url{https://github.com/funket/pysbm}, commit 4239d57. 

The following Python script was used to compute the estimated clustering under different PySBM models, specified by \verb+inner_sbm_model+, using different inference algorithms, specified by \verb+inference_method+. For \verb+num_blocks+, for the models SPC, DCPU, and DCPUH, we used the number of blocks inferred by graph-tool using the same model (NDC-Flat, DC-Flat with ``uniform'' prior, and DC-Flat with ``distributed'' prior, respectively). For the remaining models, we used the number of blocks in the Chosen-Flat clustering.

\begin{lstlisting}[language=Python]
models = {
    "SKN": pysbm.TraditionalUnnormalizedLogLikelyhood,
    "DCKN": pysbm.DegreeCorrectedUnnormalizedLogLikelyhood,
    "DCP": pysbm.DegreeCorrectedMicrocanonicalEntropy,
    "ICLexJ": pysbm.IntegratedCompleteLikelihoodExactJeffrey,
    "ICLexU": pysbm.IntegratedCompleteLikelihoodExactUniform,
    "SNR": pysbm.NewmanReinertNonDegreeCorrected,
    "DCNR": pysbm.NewmanReinertDegreeCorrected,
    "SPC": pysbm.LogLikelihoodOfFlatMicrocanonicalNonDegreeCorrected,
    "DCPU": pysbm.LogLikelihoodOfFlatMicrocanonicalDegreeCorrectedUniform,
    "DCPUH": pysbm.LogLikelihoodOfFlatMicrocanonicalDegreeCorrectedUniformHyperprior
}
methods = {
    "PAH": pysbm.PeixotoInference,
    "KL-EM": pysbm.EMInference,
    "MHA-250k": pysbm.MetropolisHastingInferenceTwoHundredFiftyK,
}

g = nx.read_edgelist(input_network, delimiter="\t", nodetype=int)
partition = pysbm.NxPartitionGraphBased(
    graph=g, number_of_blocks=num_blocks, save_degree_distributions=True
)
objective_function = models[inner_sbm_model](is_directed=false)
inference = methods[inference_method](g, objective_function, partition)
inference.infer_stochastic_block_model()
\end{lstlisting}

\clearpage

\section{Software and Commands for Treatments}
\label{suppl:software-commands/postprocess}

\verb+constrained_clustering+ is available at \url{https://github.com/MinhyukPark/constrained-clustering}, release v1.1.0.

To post-process a clustering with CC or WCC, the command is
\begin{lstlisting}
./constrained_clustering MincutOnly \
    --edgelist <network> \
    --existing-clustering <clustering> \
    --output-file <output> \
    --log-file <log> \
    --num-processors <num_processors> \
    --connectedness-criterion <criterion> \
    --log-level 1
\end{lstlisting}
where
\begin{itemize}
    \item \texttt{<network>} is the path to the edge list file of the network we want to cluster,
    \item \texttt{<clustering>} is the path to the cluster assignment file corresponding to the clustering we want to post-process,
    \item \texttt{<output>} is the path to the file storing the post-processing result,
    \item \texttt{<log>} is the path to the file storing the log of the post-processing process
    \item \texttt{<num\_processors>} is the number of CPUs to be used
    \item \texttt{criterion} is the criterion to use to determine whether or not to split the cluster (0 corresponds to CC and 1 corresponds to WCC)
\end{itemize}

\clearpage

\section{Software for Description Length Computation}
\label{suppl:dl:code}

The code is available at \url{https://github.com/illinois-or-research-analytics/network-analysis-code}, commit a92dcf8.

To compute the description length of a clustered network under a flat SBM model, the command is
\begin{lstlisting}
python compute_dl.py \
    --edgelist <network> \
    --clustering <cluster> \
    --model <model> \
    --output <output>
\end{lstlisting}
where
\begin{itemize}
    \item \texttt{<network>} is the path to the edge list file of the network,
    \item \texttt{<cluster>} is the path to the clustering of the network,
    \item \texttt{<model>} is the model we want to use (DC for DC-Flat, Non-DC for NDC-Flat, PP for PP-Flat),
    \item \texttt{<output>} is the path to the JSON file where the results will be stored.
\end{itemize}

\clearpage

\section{Software and Commands for Accuracy Evaluation}
\label{suppl:software-commands/eval-threshold}

The code is available at \url{https://github.com/illinois-or-research-analytics/network-analysis-code}, commit ac9af43.

To compute the density of all clusters in a clustered network, the command is
\begin{lstlisting}
python analysis_threshold/density.py \
    --network <network> \
    --cluster <cluster> \
    --outfile <output>
\end{lstlisting}
where
\begin{itemize}
    \item \texttt{<network>} is the path to the edge list file of the network,
    \item \texttt{<clustering>} is the path to the cluster assignment file for the network,
    \item \texttt{<output>} is the path to the output CSV file.
\end{itemize}
The output CSV file has as many lines as the number of nodes in the network. The first column is the node ID. The second column is the cluster to which the node is assigned. The third column is the density of the cluster to which the node is assigned.

To compute the accuracy at different threshold values (to filter ground-truth clusters), the command is
\begin{lstlisting}
python analysis_threshold/accuracy.py \
    --network <network> \
    --gt_cluster <density-output> \
    --et_cluster <cluster> \
    --output <output> \
    --greater True \
    --minsize 1
\end{lstlisting}
where
\begin{itemize}
    \item \texttt{<network>} is the path to the edge list file of the network,
    \item \texttt{<density-output>} is the path to the output of the previous script, which has the information of both the ground-truth clustering and the density of all ground-truth clusters,
    \item \texttt{<cluster>} is the path to the estimated clustering produced by community detection techniques (SBM in this case),
    \item \texttt{<output>} is the path to the directory containing the accuracy results.
\end{itemize}

To further assess clustering \textcolor{black}{accuracy}, we computed precision and recall, as we now describe.
Treating both the true and estimated clusterings as equivalence relations---each defined by a set of node pairs where 
$(x,y)$ belongs to the relation if and only if nodes $x$ and $y$ are in the same cluster---we define:
\begin{itemize}
    \item False negatives (FN): Pairs present in the true clustering but missing in the estimated clustering.
    \item False positives (FP): Pairs present in the estimated clustering but absent in the true clustering.
    \item True positives (TP): Pairs present in both the true and estimated clusterings.
    \item True negatives (TN): Pairs absent from both clusterings.
\end{itemize}
Precision is then $\frac{TP}{TP+FP}$ and Recall is $\frac{TP}{TP+FN}$.
\clearpage


\section{Networks for the PySBM Study}
\label{suppl:pysbm/networks}

We analyzed the $10$ smallest networks in the list, with the number of nodes ranging from $906$ to $2115$. Specifically, the networks are \texttt{dnc}; \texttt{uni\_email}; \texttt{polblogs}; \texttt{faa\_routes}; \texttt{netscience}; \texttt{new\_zealand\_collab}; \texttt{collins\_yeast}; \texttt{interactome\_stelzl}; \texttt{bible\_nouns}; \texttt{at\_migrations}.

\clearpage

\section{Additional Details for Experiment 4}

\subsection{Networks}
\label{suppl:dl:networks}

There are $65$ networks (out of $74$ studied networks) that selected DC-Flat as the Chosen-Flat model. The list includes \texttt{academia\_edu}, \texttt{advogato}, \texttt{anybeat}, \texttt{at\_migrations}, \texttt{berkstan\_web}, \texttt{bible\_nouns}, \texttt{bitcoin\_alpha}, \texttt{bitcoin\_trust}, \texttt{chess}, \texttt{citeseer}, \texttt{cora}, \texttt{dblp\_cite}, \texttt{dnc}, \texttt{douban}, \texttt{drosophila\_flybi}, \texttt{elec}, \texttt{email\_enron}, \texttt{email\_eu}, \texttt{epinions\_trust}, \texttt{faa\_routes}, \texttt{facebook\_wall}, \texttt{fediverse}, \texttt{fly\_hemibrain}, \texttt{fly\_larva}, \texttt{google}, \texttt{google\_plus}, \texttt{hyves}, \texttt{inploid}, \texttt{interactome\_figeys}, \texttt{interactome\_stelzl}, \texttt{interactome\_vidal}, \texttt{internet\_as}, \texttt{jdk}, \texttt{jung}, \texttt{lastfm\_aminer}, \texttt{libimseti}, \texttt{linux}, \texttt{livemocha}, \texttt{lkml\_reply}, \texttt{marker\_cafe}, \texttt{marvel\_universe}, \texttt{myspace\_aminer}, \texttt{new\_zealand\_collab}, \texttt{notre\_dame\_web}, \texttt{openflights}, \texttt{petster}, \texttt{pgp\_strong}, \texttt{polblogs}, \texttt{prosper}, \texttt{python\_dependency}, \texttt{reactome}, \texttt{slashdot\_threads}, \texttt{slashdot\_zoo}, \texttt{stanford\_web}, \texttt{topology}, \texttt{twitter}, \texttt{twitter\_15m}, \texttt{uni\_email}, \texttt{us\_air\_traffic}, \texttt{wiki\_link\_dyn}, \texttt{wiki\_rfa}, \texttt{wiki\_users}, \texttt{wikiconflict}, \texttt{word\_assoc}, \texttt{yahoo\_ads}.

There is only one network (out of the $65$ above) where, by removing the $- log p(e)$ component, both DC-Flat+CC and DC-Flat are equally preferred with respect to minimizing the description length. The network is \texttt{at\_migrations}.

\subsection{Codes}
\label{suppl:dl:weighted-cd-code}

To run the DC-Flat model with different weight configurations, the following code was used. Here, \texttt{weight} corresponds to $\beta$, and \texttt{use\_pe = False} corresponds to deactivating the $-\log p(e)$ component.

\begin{lstlisting}[language=python]
import graph_tool.all as gt

def run_weighted_sbmcd(edgelist_fn, weight=1.0, use_pe=True):
    g = gt.load_graph_from_csv(edgelist_fn, csv_options={"delimiter": "\t"})
    gt.remove_parallel_edges(g)
    gt.remove_self_loops(g)
    return gt.minimize_blockmodel_dl(
        g,
        state=gt.BlockState,
        state_args={
            "deg_corr": True,
        },
        multilevel_mcmc_args={
            "entropy_args": {
                "multigraph": False,
                "beta_dl": weight,
                "edges_dl": use_pe,
            },
        },
    )
\end{lstlisting}

\clearpage

\section{Additional Figures}

\begin{figure}[!htpb]
	\centering
	\includegraphics[]{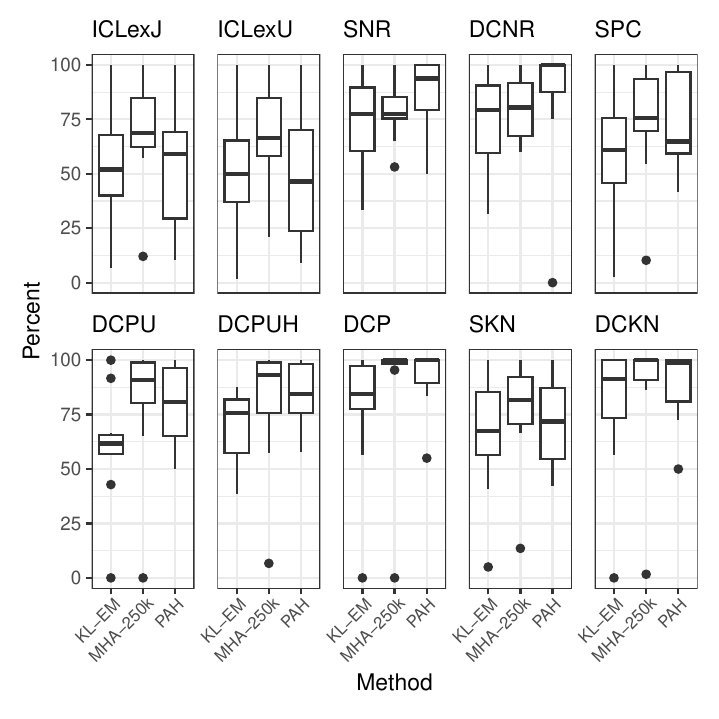}
	\caption[Experiment 1: Connectivity of clusters produced by PySBM inference algorithms on real-world networks]{
		\textbf{Experiment 1: Percentage of non-singleton  clusters produced by PySBM algorithms  on real-world networks that are internally disconnected} 
            The results are on the $10$ smallest networks.
            Each subfigure corresponds to one of ten SBM models implemented in PySBM. 
            For each model, three inference algorithms implemented in PySBM are used. 
            The KL-EM algorithm has the lowest frequency of disconnected clusters.
	}
	\label{suppl:fig:pysbm-models-conn}
\end{figure}

\clearpage



\begin{figure}[!ht]
\centering
\includegraphics[width=0.9\textwidth]{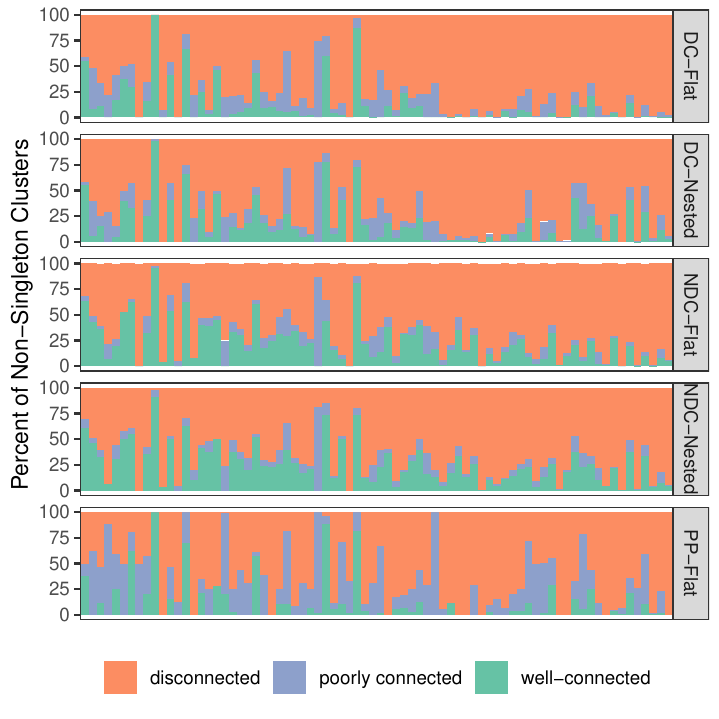}
\caption[Experiment 1: Cluster connectivity of SBM clustering on real-world non-bipartite networks]{\textbf{Experiment 1: Cluster connectivity in SBM clustering on real-world non-bipartite networks.} We show the percentage of non-singleton clusters that are disconnected (orange), poorly-connected (blue), and well-connected (green) for all graph-tool SBM models. Each bar represents one of $76$ networks ranging in size from $906$ to almost $6.3M$ nodes, and the networks are sorted in increasing number of nodes.}
\label{suppl:fig:non-bipartite-conn}
\end{figure}

\clearpage


\begin{figure}[!ht]
\centering
\includegraphics[width=0.9\textwidth]{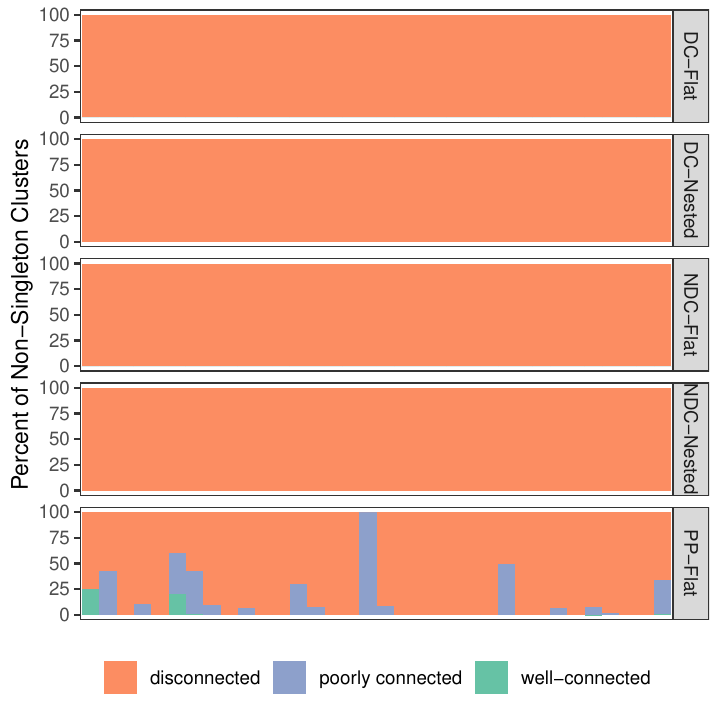}
\caption[Experiment 2: Cluster connectivity of SBM clustering on real-world bipartite networks]{\textbf{Experiment 2: Cluster connectivity of SBM clustering on real-world bipartite networks.} We show the percentage of non-singleton clusters that are disconnected (orange), poorly-connected (blue), and well-connected (green) for all graph-tool SBM models. Each bar represents one of $34$ networks ranging in size from $1884$ to $757,730$ nodes, and the networks are sorted in increasing number of nodes. }
\label{suppl:fig:bipartite-conn}
\end{figure}





\begin{figure}[!htpb]
    \centering
    \includegraphics[width=\textwidth]{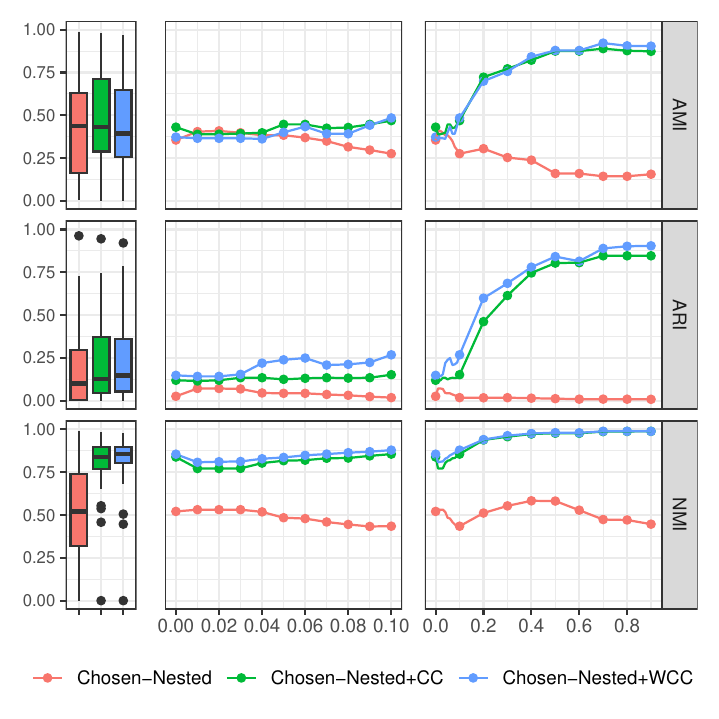}
    \caption[Experiment 3: Effect of treatments on accuracy of Chosen-Nested on EC-SBM Leiden-CPM(0.01) networks]{\textbf{Experiment 3: Effect of treatments on accuracy of Chosen-Nested on EC-SBM Leiden-CPM(0.01) networks} The results are from running Chosen-Nested on $73$ EC-SBM networks, using Leiden-CPM(0.01) for input clustering (missing \texttt{myspace\_aminer} due to memory issues for WCC). The leftmost column shows the accuracy when all clusters in each network are considered. For the middle and rightmost columns, the $x$-axis represents a density threshold. In these columns, only clusters with a density strictly greater than the threshold are analyzed. 
    Here, singleton clusters are included in the $0.0$ threshold.
    The middle column focuses specifically on including the low-density clusters. The plotted value is the median accuracy across all networks for each threshold.}
    \label{suppl:fig:nested-leiden-cpm-0.01}
\end{figure}

\begin{figure}[!htpb]
    \centering
    \includegraphics[width=\textwidth]{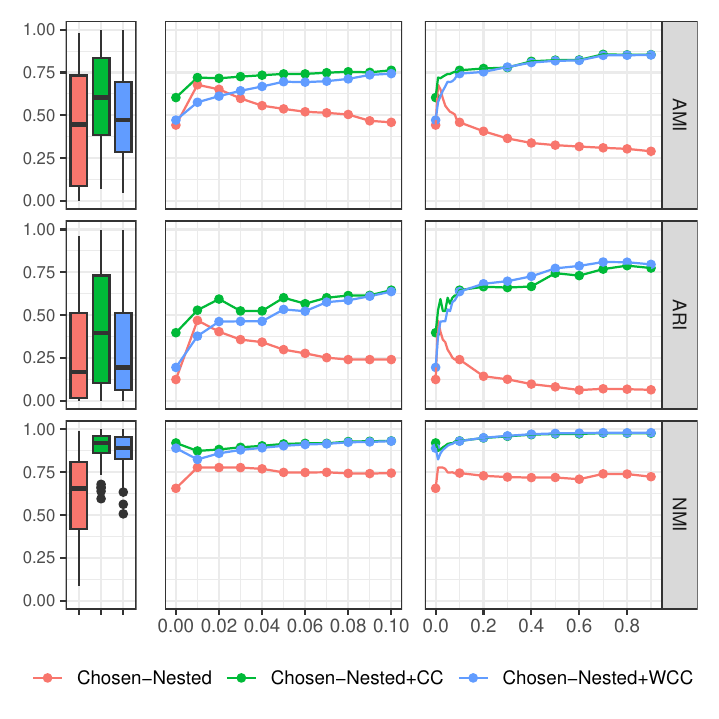}
    \caption[Experiment 3: Effect of treatments on accuracy of Chosen-Nested on EC-SBM SBM+CC networks]{\textbf{Experiment 3: Effect of treatments on accuracy of Chosen-Nested on EC-SBM SBM+CC networks} The results are from running Chosen-Nested on $74$ EC-SBM networks, using SBM+CC for input clustering. The leftmost column shows the accuracy when all clusters in each network are considered. For the middle and rightmost columns, the $x$-axis represents a density threshold. In these columns, only clusters with a density strictly greater than the threshold are analyzed. 
    Here, singleton clusters are included in the $0.0$ threshold.
    The middle column focuses specifically on including the low-density clusters. The plotted value is the median accuracy across all networks for each threshold.}
    \label{suppl:fig:nested-sbm+cc}
\end{figure}



\begin{figure}[!htpb]
    \centering
    \includegraphics[width=\textwidth]{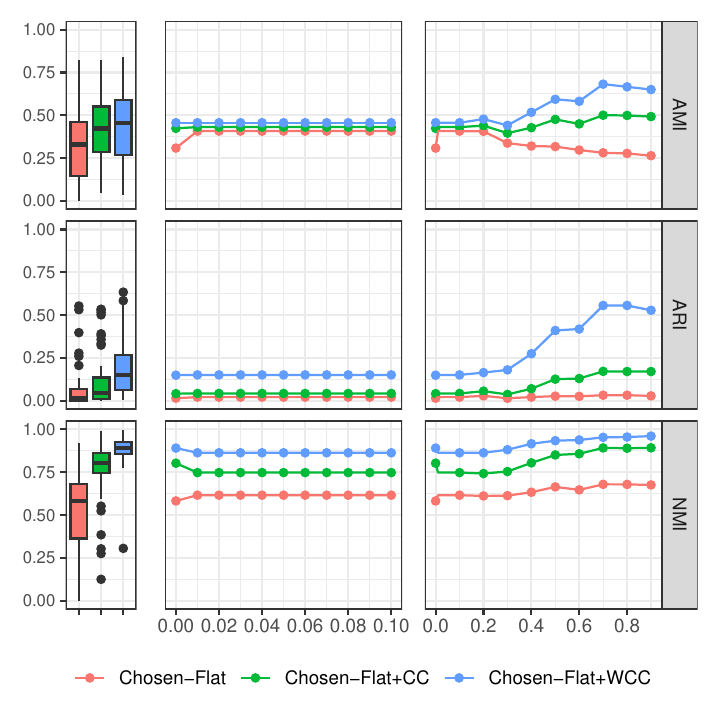}
    \caption[Experiment 3: Effect of treatments on accuracy of Chosen-Flat on EC-SBM Leiden-CPM(0.1) networks]{\textbf{Experiment 3: Effect of treatments on accuracy of Chosen-Flat on EC-SBM Leiden-CPM(0.1) networks} The results are from running Chosen-Flat on $73$ EC-SBM networks using Leiden-CPM(0.1) for input clustering (missing \texttt{myspace\_aminer} due to memory issues for WCC). Each row corresponds to an accuracy metric. The leftmost column shows the accuracy when all clusters in each network are considered. For the middle and rightmost columns, the $x$-axis represents a density threshold. In these columns, only clusters with a density strictly greater than the threshold are analyzed. 
    Here, singleton clusters are included in the $0.0$ threshold.
    The middle column focuses specifically on including the low-density clusters. The plotted value is the median accuracy across all networks for each threshold.}
    \label{suppl:fig:flat-leiden-cpm-0.1}
\end{figure}

\begin{figure}[!htpb]
    \centering
    \includegraphics[width=\textwidth]{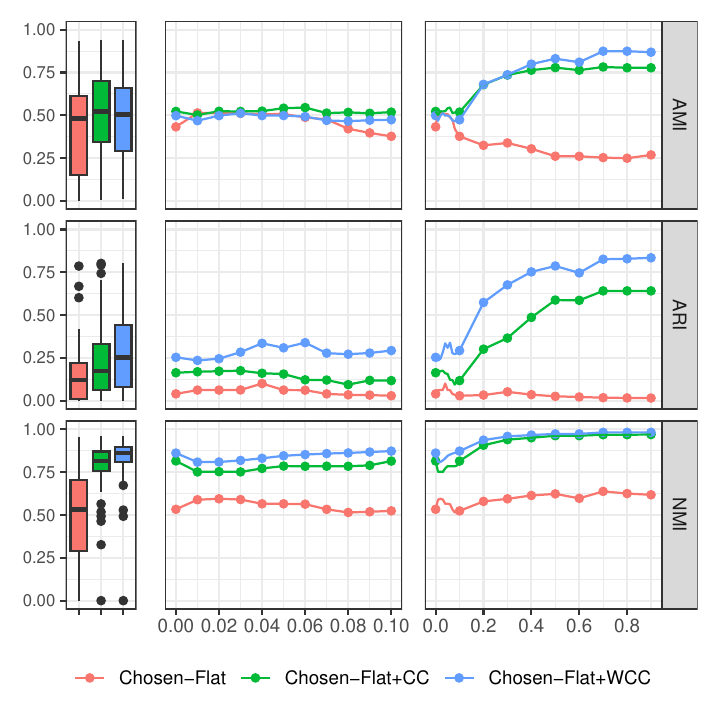}
    \caption[Experiment 3: Effect of treatments on accuracy of Chosen-Flat on EC-SBM Leiden-CPM(0.01) networks]{\textbf{Experiment 3: Effect of treatments on accuracy of Chosen-Flat on EC-SBM Leiden-CPM(0.01) networks} The results are from running Chosen-Flat on $73$ EC-SBM networks using Leiden-CPM(0.01) for input clustering (missing \texttt{myspace\_aminer} due to memory issues for WCC). The leftmost column shows the accuracy when all clusters in each network are considered. For the middle and rightmost columns, the $x$-axis represents a density threshold. In these columns, only clusters with a density strictly greater than the threshold are analyzed. 
    Here, singleton clusters are included in the $0.0$ threshold.
    The middle column focuses specifically on including the low-density clusters. The plotted value is the median accuracy across all networks for each threshold.}
    \label{suppl:fig:flat-leiden-cpm-0.01}
\end{figure}

\begin{figure}[!htpb]
    \centering
    \includegraphics[width=\textwidth]{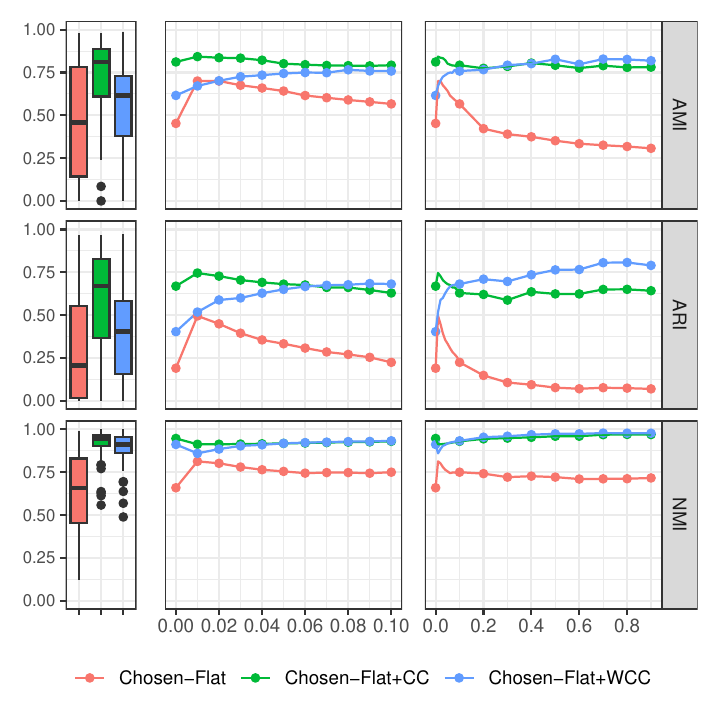}
    \caption[Experiment 3: Effect of treatments on accuracy of Chosen-Flat on EC-SBM SBM+CC networks]{\textbf{Experiment 3: Effect of treatments on accuracy of Chosen-Flat on EC-SBM SBM+CC networks} The results are from running Chosen-Flat on $74$ EC-SBM networks, using SBM+CC for input clustering. The leftmost column shows the accuracy when all clusters in each network are considered. For the middle and rightmost columns, the $x$-axis represents a density threshold. In these columns, only clusters with a density strictly greater than the threshold are analyzed. 
    Here, singleton clusters are included in the $0.0$ threshold.
    The middle column focuses specifically on including the low-density clusters. The plotted value is the median accuracy across all networks for each threshold.}
    \label{suppl:fig:flat-sbm+cc}
\end{figure}

\begin{figure}[!htpb]
    \centering
    \includegraphics[width=\textwidth]{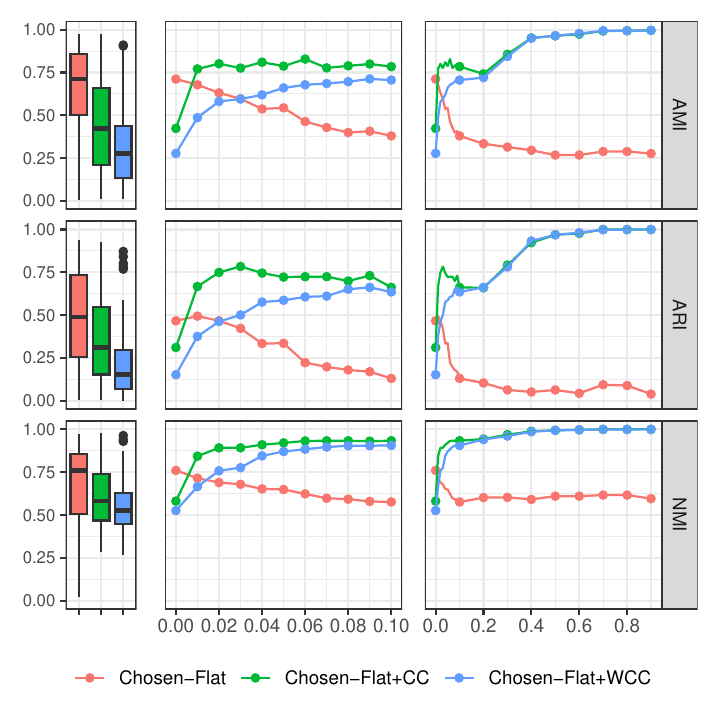}
    \caption[Experiment 3: Effect of treatments on accuracy of Chosen-Flat on EC-SBM Leiden-Mod networks]{\textbf{Experiment 3: Effect of treatments on accuracy of Chosen-Flat on EC-SBM Leiden-Mod networks} The results are from running Chosen-Flat on $73$ EC-SBM networks, using Leiden-Mod for input clustering (missing \texttt{petster} due to time limit for WCC). The leftmost column shows the accuracy when all clusters in each network are considered. For the middle and rightmost columns, the $x$-axis represents a density threshold. In these columns, only clusters with a density strictly greater than the threshold are analyzed. 
    Here, singleton clusters are included in the $0.0$ threshold.
    The middle column focuses specifically on including the low-density clusters. The plotted value is the median accuracy across all networks for each threshold.}
    \label{suppl:fig:flat-leiden-mod}
\end{figure}


\begin{figure}[!htpb]
    \centering
    \includegraphics[width=\textwidth]{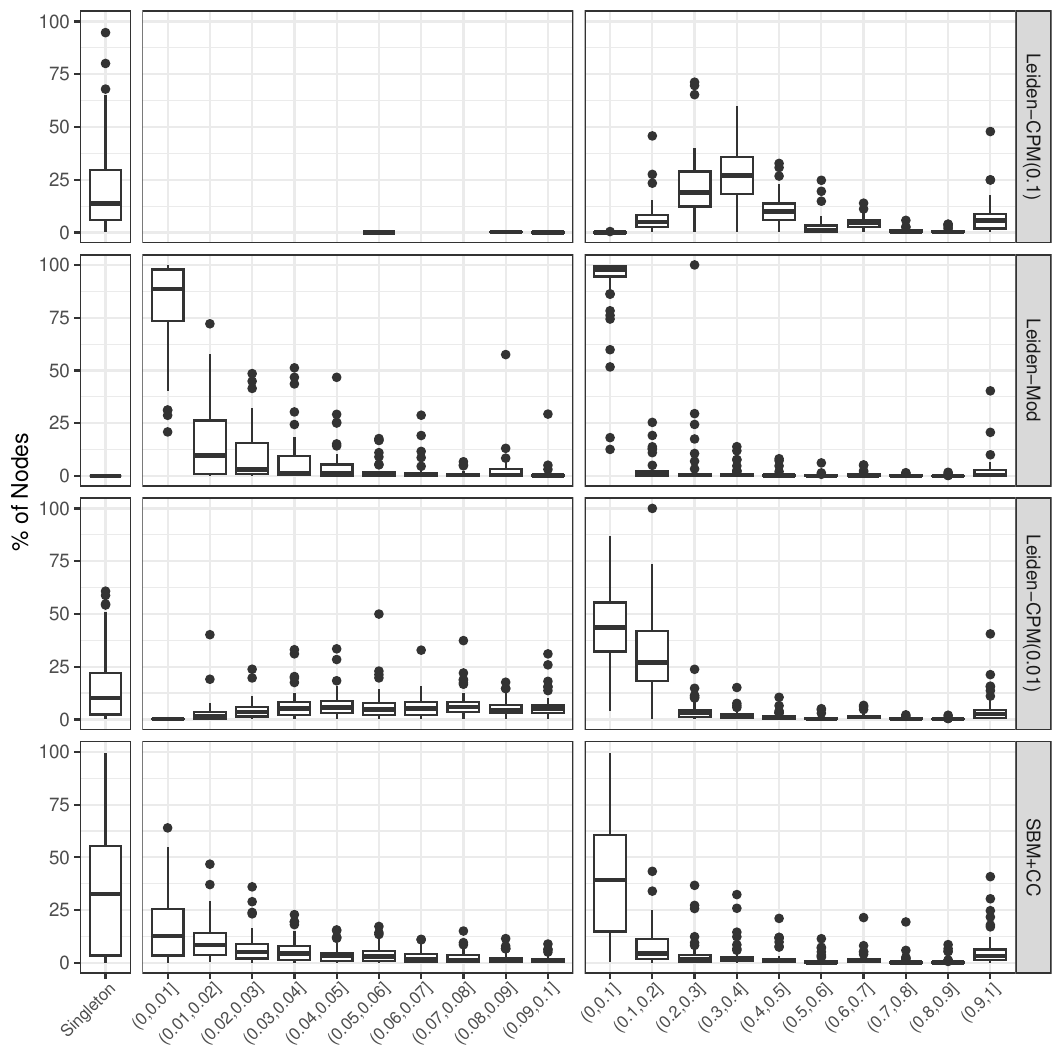}
    \caption[Experiment 3: Node coverage of EC-SBM ground-truth clusterings]{\textbf{Experiment 3: Node coverage of EC-SBM ground-truth clusterings} Each row corresponds to a different set of $74$ networks. The rows are distinguished by the input clustering method specified in the row label. The $x$-axis indicates a density bin. 
    No boxes indicate that there are no nodes in that bin.
    }
    \label{suppl:fig:node-coverage-density}
\end{figure}

\begin{figure}[!htpb]
    \centering
    \includegraphics[width=\textwidth]{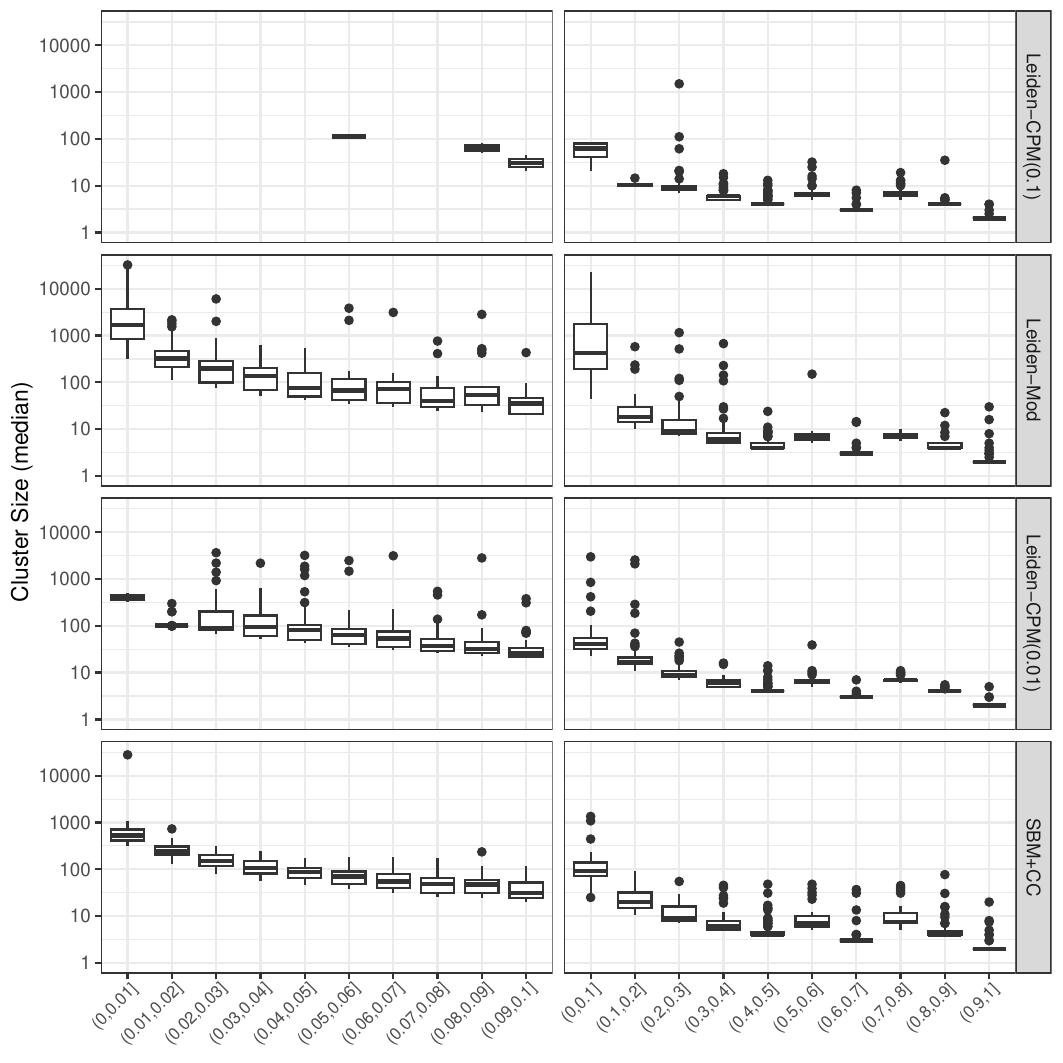}
    \caption[Experiment 3: Cluster sizes of EC-SBM ground-truth clusterings]{\textbf{Experiment 3: Cluster sizes of EC-SBM ground-truth clusterings} Each row corresponds to a different set of $74$ networks. The rows are distinguished by the input clustering method specified in the row label. The $x$-axis indicates a density bin. For each network, we compute the median cluster size (excluding singletons) in the network.
    }
    \label{suppl:fig:cluster-size-median-density}
\end{figure}

\begin{figure}
   \centering
    \includegraphics[width=\textwidth]{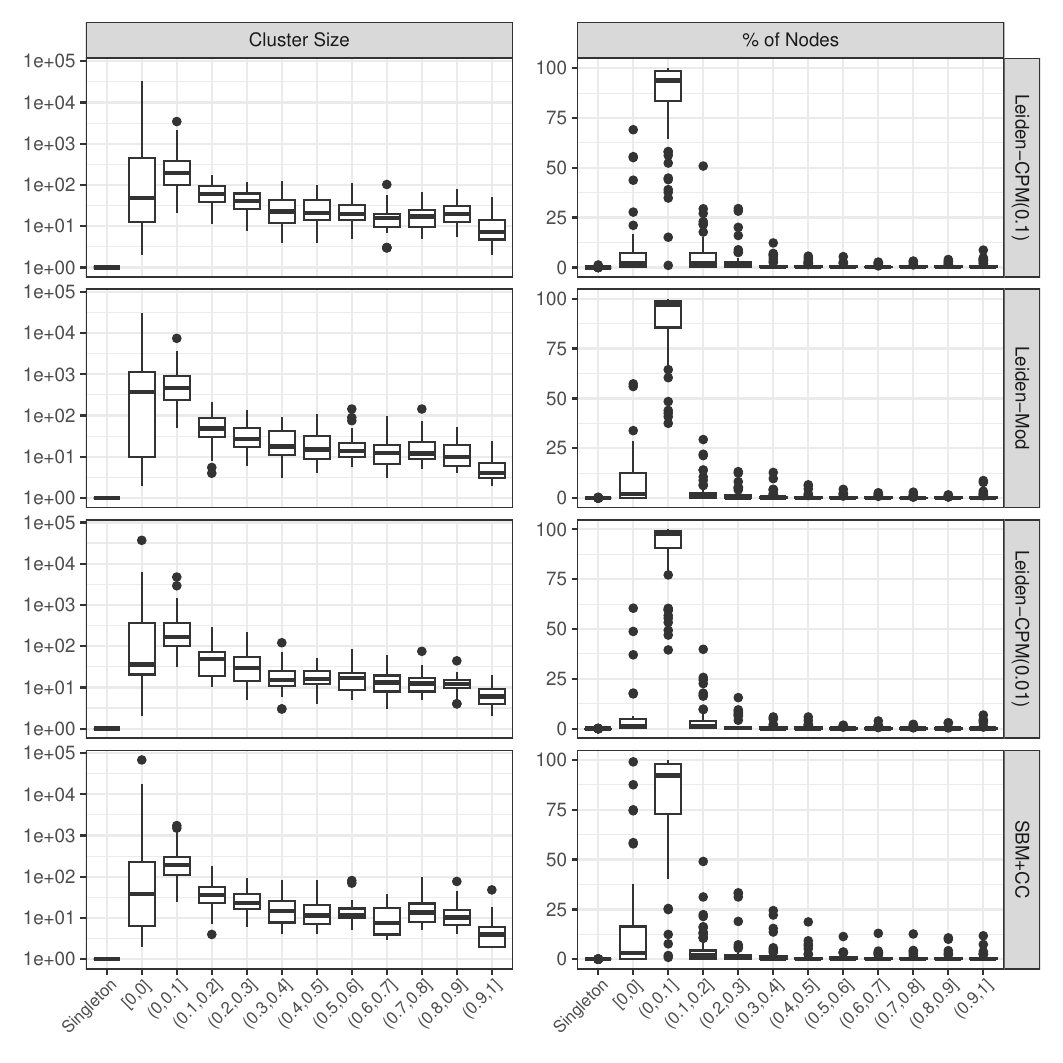}
   \caption[Experiment 3: Statistics for Chosen-Flat clusterings on synthetic networks]{\textbf{Experiment 3: Statistics for Chosen-Flat clusterings on Synthetic Networks.}  Each row indicates a collection of 74 EC-SBM networks that differ by the input clustering; the left column shows cluster sizes and the right column shows percentage of nodes, binned by cluster density range. For all synthetic networks, there are no singleton clusters, but many clusters have no edges (indicated by density $0.0$), and more than 90\% of the nodes are in very sparse clusters (density at most $0.1$). Also note that there is a small range of cluster sizes, but also some very small clusters (below size 10). }
   \label{fig:suppl-stats-chosenflat-synhetic}
\end{figure}


\begin{figure}[!htpb]
    \centering
    \includegraphics[width=\textwidth]{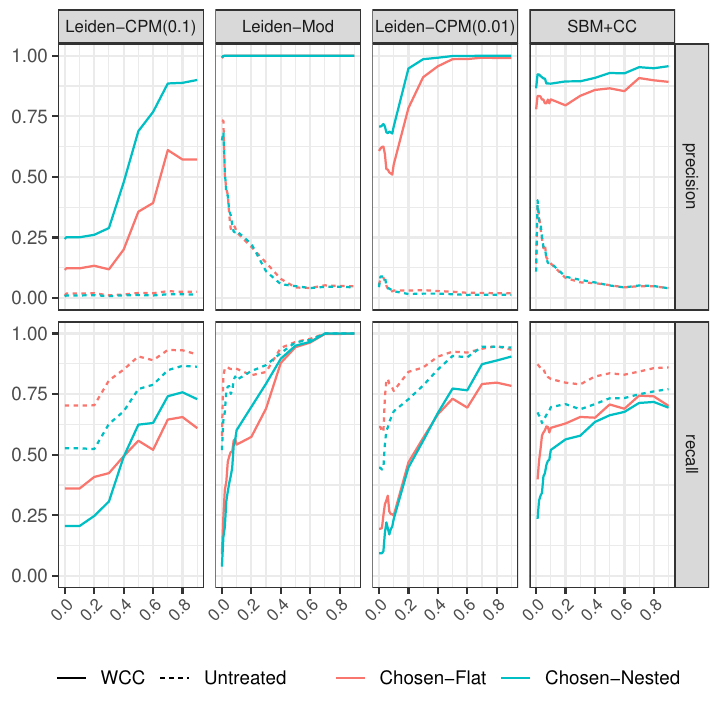}
    \caption[Experiment 3: Precision and recall for treated and untreated SBM clusterings]{\textbf{Experiment 3: Precision and recall for treated and untreated SBM clusterings}  
    Each column corresponds to a different set of EC-SBM networks, distinguished by the input clustering method specified in the column label. Specifically, from left to right, there are $73, 71, 73,$ and $74$ networks (missing \texttt{myspace\_aminer} for Leiden-CPM at both resolutions and additionally \texttt{berkstan\_web} and \texttt{petster} for Leiden-Mod, due to memory issues for WCC on these networks).
    Each row corresponds to an accuracy metric and values shown are medians across all networks for that threshold. The $x$-axis indicates a density threshold. For each threshold value, only clusters with a density strictly greater than the threshold value are analyzed. 
    Here, singleton clusters are included in the $0.0$ threshold, and when TP = FP = 0, we say precision is 1.0.
    We see that for all methods, recall increases as density increases. However, the impact on precision differs: for untreated Chosen-Flat and Chosen-Nested, precision tends to decrease as density increases, while precision tends to increase as density increases for  WCC-treated 
    Chosen-Flat and Chosen-Nested.
    \label{suppl:figs:prec-recall-density}
    }
\end{figure}

\clearpage

\bibliography{clustering}